\documentclass[aps,prl,floatfix,twocolumn,superscriptaddress,longbibliography]{revtex4-2}
\usepackage{amsmath}
\usepackage{hyperref}
\usepackage{amssymb }
\usepackage{color}
\usepackage{graphicx}
\usepackage{bm}
\usepackage{multirow}
\usepackage{setspace}
\usepackage[normalem]{ulem}

\begin{document}
\title{Quadrupolar magnetic excitations in an isotropic spin-1 antiferromagnet}

\author{A. Nag}\email[]{abhishek.nag@diamond.ac.uk}
\affiliation{Diamond Light Source, Harwell Campus, Didcot OX11 0DE, United Kingdom}

\author{A. Nocera}\email[]{alberto.nocera@ubc.ca}
\affiliation{Stewart Blusson Quantum Matter Institute, University of British Columbia, Vancouver, British Columbia, V6T 1Z4 Canada}
\affiliation{Department of Physics Astronomy, University of British Columbia, Vancouver, British Columbia, Canada V6T 1Z1}

\author{S. Agrestini}
\author{M. Garcia-Fernandez}
\author{A. C. Walters}
\affiliation{Diamond Light Source, Harwell Campus, Didcot OX11 0DE, United Kingdom}

\author{Sang-Wook Cheong}
\affiliation{Rutgers Center for Emergent Materials, Rutgers University, Piscataway, NJ, USA.}

\author{S. Johnston}\email[]{sjohn145@utk.edu}
\affiliation{Department of Physics and Astronomy, The University of Tennessee, Knoxville, Tennessee 37966, USA}

\author{Ke-Jin Zhou}\email[]{kejin.zhou@diamond.ac.uk}
\affiliation{Diamond Light Source, Harwell Campus, Didcot OX11 0DE, United Kingdom}

\begin{abstract}
The microscopic origins of emergent behaviours in condensed matter systems are encoded in their excitations. In ordinary magnetic materials, single spin-flips give rise to collective dipolar magnetic excitations called magnons. Likewise, multiple spin-flips can give rise to multipolar magnetic excitations in magnetic materials with spin $S \ge 1$. Unfortunately, since most experimental probes are governed by dipolar selection rules, collective multipolar excitations have generally remained elusive. For instance, only dipolar magnetic excitations have been observed in isotropic $S=1$ Haldane spin systems. Here, we unveil a hidden quadrupolar constituent of the spin dynamics in antiferromagnetic $S=1$ Haldane chain material Y$_{2}$BaNiO$_{5}$ using Ni ${L_3}$-edge resonant inelastic x-ray scattering. Our results demonstrate that pure quadrupolar magnetic excitations can be probed without direct interactions with dipolar excitations or anisotropic perturbations. Originating from on-site double spin-flip processes, the quadrupolar magnetic excitations in Y$_{2}$BaNiO$_{5}$ show a remarkable dual nature of collective dispersion. While one component propagates as non-interacting entities, the other behaves as a bound quadrupolar magnetic wave. This result highlights the rich and largely unexplored physics of higher-order magnetic excitations.
\end{abstract}

\flushbottom
\maketitle

\section*{Introduction}
The elementary excitation of a magnetically ordered material is a single dipolar spin-flip of an electron, delocalised coherently across the system in the form of a spin-wave. The spin-wave quasiparticle, known as a magnon, carries a spin angular momentum of one unit and has well-defined experimental signatures. Dipolar collective excitations also appear in low-dimensional magnets that remain disordered to the lowest achievable temperatures because of quantum fluctuations. A paradigmatic example is the $S=1$ antiferromagnetic Haldane spin chain, where magnetic order is suppressed in favour of a singlet ground state with non-local topological order~\cite{haldane1983pla,kennedy1992cmp}. Several theoretical and experimental works have established that a single spin-flip from this exotic ground state creates dipolar magnetic excitations that propagate along the chain above an energy gap of $~\Delta_{\mathrm{H}}\sim0.41 J$, the Haldane gap~\cite{kenzelmann2002prb1,zaliznyak2001prl,xu2007sci,xu1996prb,white2008prb,affleck1988cmp}.
%,binder2020prb,

\begin{figure*}[ht]
	\centering
	\includegraphics[width=1\textwidth]{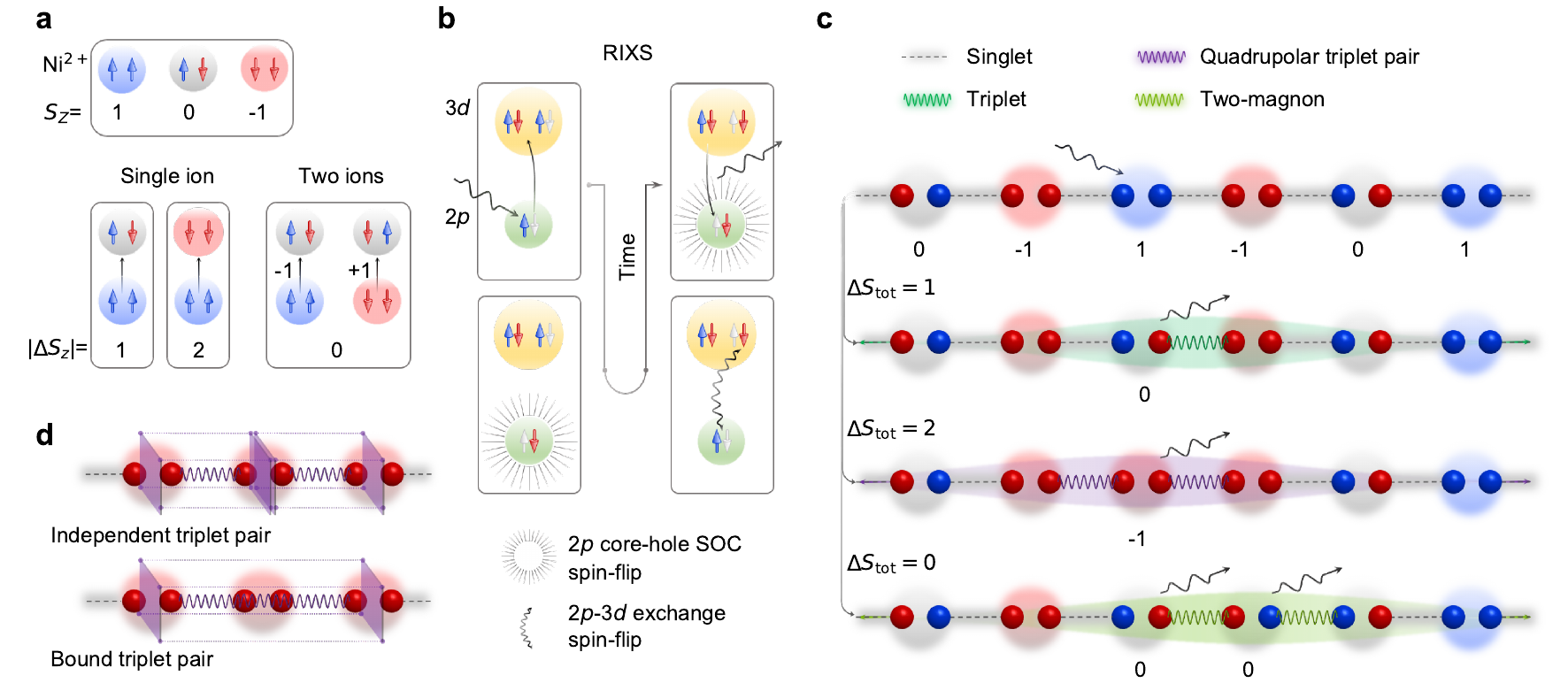}
	\caption{\textbf{Magnetic excitations in $\mathbf{S=1}$ Haldane  chains.}
		\textbf{a,} Possible unpaired spin configurations in an $S=1$ ion (Ni$^{2+}$) and spin-flip transitions. \textbf{b,} Process of probing single-site quadrupolar  $\Delta S_\mathrm{tot}=2$ excitations in $L$-edge RIXS. An $L$-edge RIXS process of a $S=1$ 3$d^8$ ion (with ground state with $S_\mathrm{tot}=0$) proceeds via creation of an intermediate 2$p^5$ 3$d^9$ state. In this intermediate state, due to the strong spin-orbit coupling of 2$p^5$ core-hole, the spin-angular momentum is not conserved leading to a valence orbital spin-flipped final state with $S_\mathrm{tot}=1$. Additionally, due to the many-body core-valence Coulomb exchange interactions in the intermediate states, an additional available valence orbital spin-flip may occur thereby creating an excited final state with $S_\mathrm{tot}=2$~\cite{nag2020prl}. \textbf{c,} A representative Haldane spin chain characterised by the topological non-local (string) order of alternating $S_z=\pm1$ sites intervened by any number of $S_z=0$ sites. Between neighbouring sites a pair of spin-1/2s form an antisymmetric singlet. Single-site single spin-flips give rise to dipolar singlet-triplet excitations while single-site double spin-flips give rise to quadrupolar excitations resulting in pairs of triplets. Single spin-flips at multiple sites give rise to two-magnons. \textbf{d,} Schematic representation of dual nature of propagation of the pair of triplets formed after quadrupolar excitation.
	}
	\label{fig:1}
\end{figure*}

Materials hosting $3d$ transition metal ions with a $d^2$ or $d^8$ configuration (such as Ni$^{2+}$ for the latter) often possess strongly interacting spin-1 local magnetic moments with quantised spin projections $S_z=-1,0,1$.  In addition to the usual single spin-flip excitations, it is possible to create quadrupolar excitations by changing the composite spin by two units. Such excitations can be conceived as flipping two of the constituent spin-1/2's, as shown in Fig.~\ref{fig:1}a. 
Incidentally, quadrupolar magnetic waves arising from such transitions and carrying two units of angular momentum were predicted for $S=1$ ferromagnetic chains as early as the 1970s~\cite{tsao1975prb,sivardiere1975jmmm} and may play a role in the iron pnictide superconductors~\cite{liu2020prb}.
Since most probes are restricted by dipolar selection rules, however, such quadrupolar excitations have largely evaded detection except in rare situations where they are perturbed by anisotropic interactions, spin-orbit coupling, lattice vibrations, or large magnetic fields~\cite{akaki2017prb, ward2017prl, bai2021np, kohama2019pnas, dally2020prl, patri2019nc}. Quadrupolar magnetic excitations have never been observed in isotropic Haldane spin chains, even though dipolar magnetic excitations have been extensively studied~\cite{kenzelmann2002prb1, zaliznyak2001prl, xu2007sci, xu1996prb, white2008prb, binder2020prb, affleck1988cmp}. 
It is then natural to wonder whether purely quadrupolar collective excitations exist in isotropic spin-1 systems. 

Here, we uncover the presence of collective quadrupolar magnetic excitations in the \emph{isotropic} $S=1$ Haldane chain system Y$_2$BaNiO$_5$ using high energy-resolution Ni $L_3$-edge resonant inelastic x-ray scattering (RIXS). Previous studies on nickelates have already shown that Ni $L_3$-edge RIXS can probe dipolar magnons~\cite{fabbris2017prl,lin2021prl}. In addition, Ref.~\cite{nag2020prl} recently showed that double spin-flips are allowed at this edge through the combined many-body effect of core-valence exchange and core-hole spin-orbit interactions~\cite{groot1998prb,haverkort2010prl} (Fig.~\ref{fig:1}b), making it the optimal tool for this study. Y$_2$BaNiO$_5$ is one of the best realisations of the isotropic Haldane spin chain material with intra-chain exchange $J\sim24$~meV and negligibly small single-ion anisotropy $\sim0.035J$, exchange-anisotropy $\sim0.011J$ and inter-chain exchange $\sim0.0005J$~\cite{xu1996prb,sakaguchi1996jpsj}.  This aspect allows us to describe the system completely using a simple Heisenberg model, and emphasises the relevance of the pure quadrupolar magnetic excitations in the spin dynamics of $S=1$ systems. 

\section*{Excitations in Haldane spin chains}
In a valence bond singlet (VBS) scheme, the Heisenberg model's ground state in the Haldane phase can be represented as a macroscopic $S_\mathrm{tot}=0$ state  comprised of pairs of fictitious spin-$\frac{1}{2}$ particles on neighbouring sites that form antisymmetric singlets on each bond~\cite{affleck1988cmp} (Fig.~\ref{fig:1}c). A single local spin-flip breaks a bond singlet to form a triplet excitation, raising the chain's total spin quantum number to $S_\mathrm{tot}=1$. In contrast, a local double spin-flip would disrupt singlets on either side of the excited site, creating a pair of triplet excitations and raising the spin quantum number to $S_\mathrm{tot}=2$. In the RIXS process, it is also possible to have a total spin-conserved two-site excitation with $\Delta S_\mathrm{tot}=0$ (see Fig.~\ref{fig:1}c), which appears as a two-magnon continuum. To simplify our notation, we will refer to the single-site single spin-flip induced dipolar $\Delta S_\mathrm{tot}=1$  excitation as $\Delta S_1$, the single-site double spin-flip induced  quadrupolar $\Delta S_\mathrm{tot}=2$ excitation as $\Delta S_2$, and the two-site $\Delta S_\mathrm{tot}=0$ two-magnon excitation as $\Delta S_0$.

\begin{figure}[ht]
	\centering
	\includegraphics[width=1\linewidth]{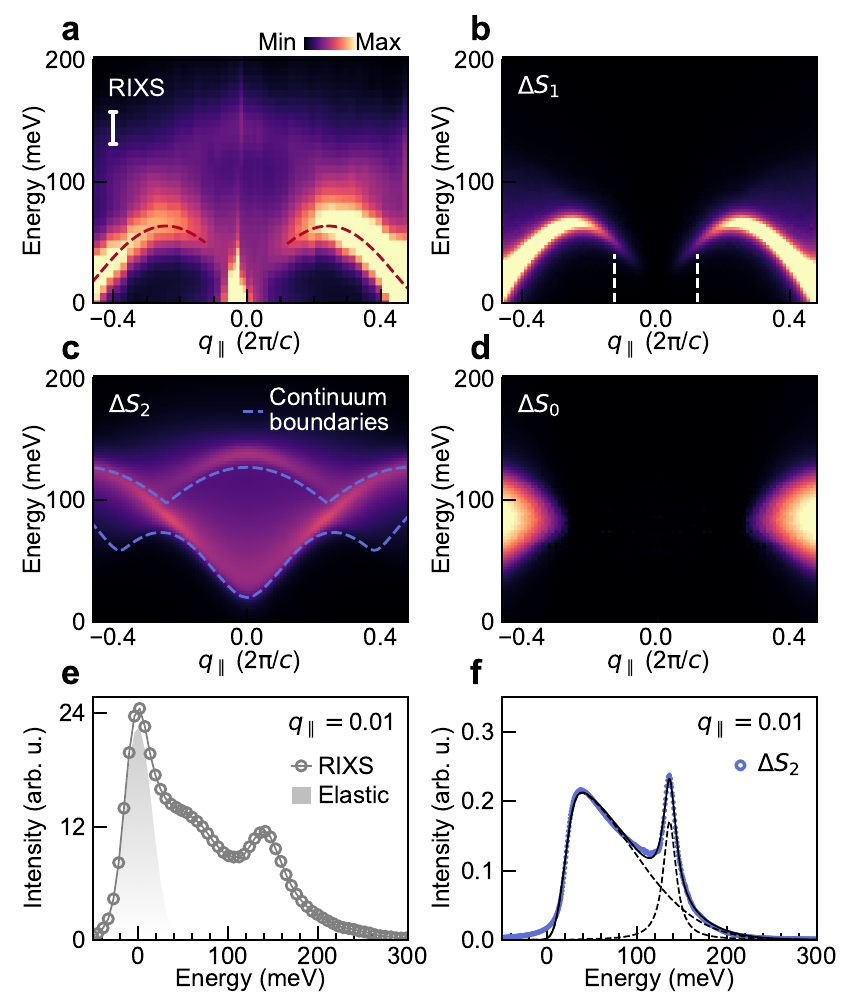}
	\caption{\textbf{RIXS results and DMRG calculations for Y$_2$BaNiO$_5$.}
		\textbf{a,} Experimental RIXS intensity map at Ni $L_3$-edge at 11~K. Dashed line is a semi-quantitative dispersion $\omega^2(q_{\parallel})=\Delta^2_\mathrm{H}+v^2\mathrm{sin}^2q_{\parallel}+\alpha^2\mathrm{cos}^2\frac{q_{\parallel}}{2}$ with $J=24$~meV, $\Delta_\mathrm{H}=0.41J$, $v=2.55J$ and $\alpha=1.1J$~\cite{Santos1989,xu1996prb,zaliznyak2001prl,xu2007sci}. The vertical bar shows the experimental energy resolution. DMRG calculated dynamic spin susceptibility intensity maps for the \textbf{b,} $\Delta S_1$, \textbf{c,} $\Delta S_2$ and \textbf{d,} $\Delta S_0$ excitations (see Methods for details). In panel \textbf{c} we show the continuum boundaries expected for two independent triplet excitations (equivalent to two-magnon continuum boundaries). \textbf{e,} RIXS line profile at $q_{\parallel}=0.01$ and a shaded elastic peak profile. \textbf{f,} Calculated profile for $\Delta S_2$ at $q_{\parallel}=0.01$ from DMRG. The $\Delta S_2$ spectral weight can be decomposed into a broad continuum and a sharp peak (see Methods).  
	}
	\label{fig:2}
\end{figure}

Figure~\ref{fig:2}a shows RIXS intensity map collected on Y$_2$BaNiO$_5$ (see Methods). The feature with the largest spectral weight follows a dispersion relation consistent with inelastic neutron scattering (INS) results for the dipolar $\Delta S_1$ excitation~\cite{sakaguchi1996jpsj,zaliznyak2001prl,xu2007sci}. This feature is also reproduced in our density matrix renormalisation group (DMRG) calculations of the dynamical structure factor $S_{1}(q_{\parallel}, \omega)$ for an isotropic Heisenberg model (see Fig.~\ref{fig:2}b and Methods). 
Ni $L_3$-edge RIXS is unable to reach the  exact antiferromagnetic zone center ($q_{\parallel}=0.5$, in units of $2\pi/c$ throughout), where the Haldane gap of $\sim8.5$~meV exists. But it does probe an equally interesting region 
%of the reciprocal space 
close to $q_{\parallel}=0$. Prior work~\cite{white2008prb,zaliznyak2001prl,stone2006nat} has focused on observing the breakdown of the well-defined $\Delta S_1$ quasiparticle into a two-magnon continuum for $q_{\parallel}\lessapprox 0.12$ and the spectral weight vanishing as $q_{\parallel}^2$. Although we notice a reduction in intensity of the $\Delta S_1$ excitation in our experiment, we also observe significant inelastic spectral weight near zero energy close to $q_{\parallel}=0$ (also see Fig.~\ref{fig:2}e for the line spectrum at $q_{\parallel}=0.01$). Interestingly, a new dispersing excitation is clearly visible with an energy maximum of $\sim136$~meV at $q_\parallel = 0$. 
Of the calculated $S_{0}(q_{\parallel}, \omega)$, $S_1(q_{\parallel}, \omega)$, $S_{2}(q_{\parallel}, \omega)$ dynamical structure factors, shown in Figs.~\ref{fig:2}d, ~\ref{fig:2}b, and ~\ref{fig:2}c, respectively, only   
$S_{2}(q_{\parallel}, \omega)$ has spectral weight in this region. 
%Neither the  $S_{0}(q_{\parallel}, \omega)$ nor the $S_{1}(q_{\parallel}, \omega)$ have spectral weight close to $q_{\parallel}=0$, while $S_{2}(q_{\parallel}, \omega)$ does. 
Moreover, the $S_{2}(q_{\parallel}\sim0, \omega)$ line profile has two-components: a broad low-energy continuum and a sharp high-energy peak, as shown in Fig.~\ref{fig:2}f. It therefore appears that both the low- and high-energy spectral components seen in the experiment can be described by the quadrupolar $\Delta S_2$ excitation in this region of momentum space.

\begin{figure*}[ht]
	\centering
	\includegraphics[width=1\linewidth]{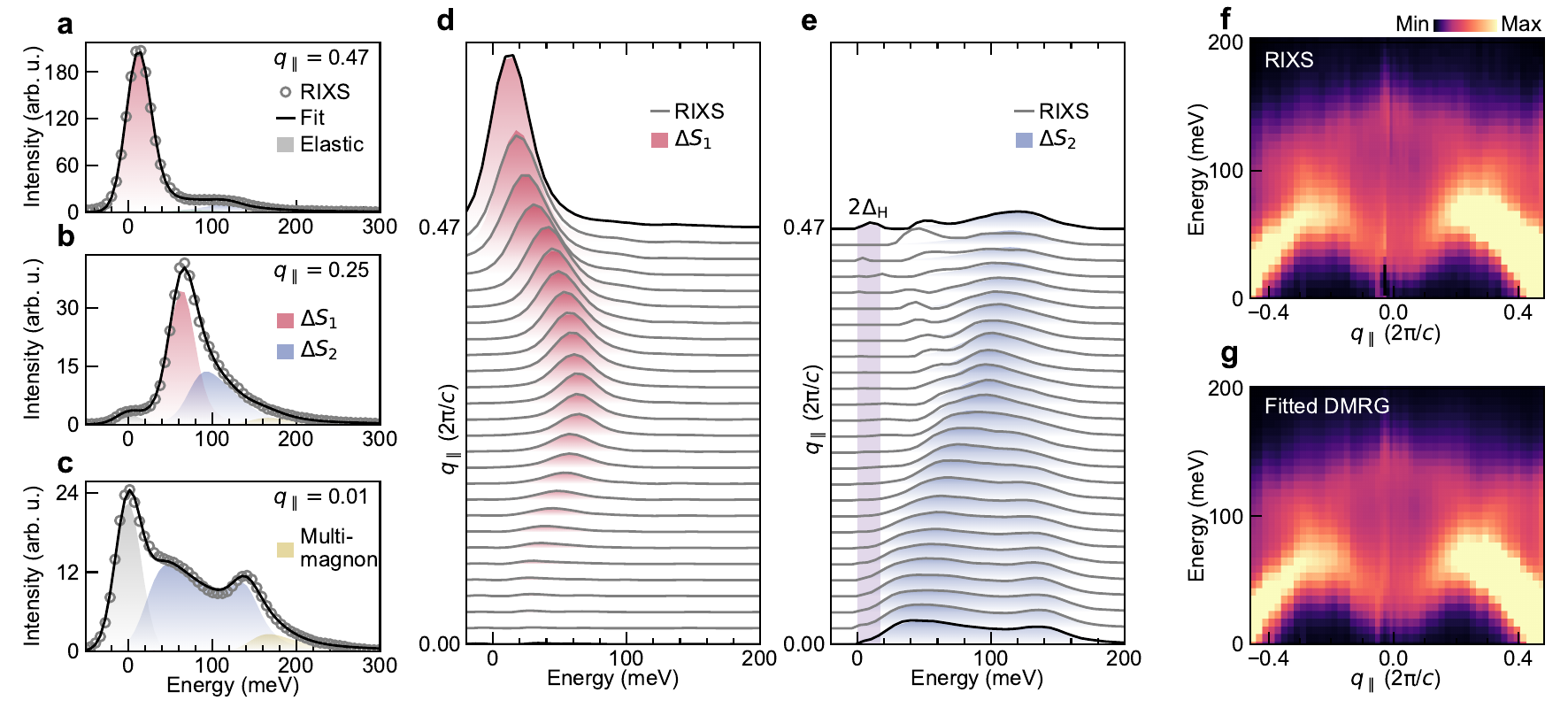}
	\caption{\textbf{Magnetic excitations probed by RIXS in Y$_2$BaNiO$_5$  at Ni $L_3$-edge.} Representative RIXS spectra at $q_{\parallel}$ = \textbf{a,} 0.47,  \textbf{b,} 0.25, and \textbf{c,} 0.01 with fits using spin susceptibilities obtained from DMRG (see Methods). 
		Note that an additional peak is included in the fits in panel \textbf{c} to account for a multi-magnon spectral tail around 200~meV, which may originate from higher-order contributions not considered in the expansion of the RIXS response. \textbf{d,} RIXS line spectra after subtraction of elastic, $\Delta S_2$ and the high energy tail structure visible in panel \textbf{c}. Shaded areas are fitted profiles from DMRG for $\Delta S_1$. \textbf{e,} RIXS line spectra after subtraction of elastic, $\Delta S_1$ and the high energy tail structure visible in panel \textbf{c}. Shaded areas are fitted profiles from DMRG for $\Delta S_2$. Vertical shaded bar represents twice the $\Delta_H$ value from Ref.~\cite{xu2007sci}. \textbf{f,} Experimental RIXS intensity map with elastic contribution subtracted. \textbf{g,} DMRG intensity map with summed contributions from  $\Delta S_0$, $\Delta S_1$ and $\Delta S_2$ excitations determined by fitting experimental RIXS spectra using Eq.~(\ref{eqn:irixs}). 
	}
	\label{fig:3}
\end{figure*}

To evaluate the relative contributions of each excitation, we decomposed the RIXS spectra across the momentum space into the dynamical structure factors calculated by DMRG. Fig.~\ref{fig:3}a-c show representative RIXS spectra.
%at $q_{\parallel}= 0.47$, $0.25$, and $0.01$, respectively. 
The data are fitted with an elastic peak and experimental energy-resolution convoluted line profiles of $S_\alpha(q_{\parallel}, \omega)$ ($\alpha = 0,1,2$, see Methods). While the spectra are dominated by the dipolar $\Delta S_1$ excitations at high $q_{\parallel}$ values, the spectral weight close to $q_{\parallel}=0$ can be fitted with only the quadrupolar $\Delta S_2$ excitation. 
%% Steve: I move this to the caption.
%(An additional peak is included in the fits to account for a spectral tail around 200~meV, which may originate from higher-order contributions not considered in the expansion of the RIXS response.)  
Fig.~\ref{fig:3}d and e show the RIXS spectra after subtracting other fitted components to keep only the $\Delta S_1$ and $\Delta S_2$ contributions, respectively, along with their fitted profiles from DMRG. This analysis shows that the RIXS spectral weight is carried by the low-energy continuum component of the $\Delta S_2$ excitations below the quasiparticle decay threshold momentum. This component is gapped and has a peak energy of $\sim19$~meV  at $q_{\parallel}~\sim0$, reminiscent of the lower boundary of the two-magnon continuum at $2\Delta_\mathrm{H}$ predicted for Haldane spin chains~\cite{white2008prb}. We note that the energy gap at this momentum has not yet been confirmed for any Haldane spin chain by INS due to the small scattering cross-sections. Fig.~\ref{fig:3}f shows the RIXS intensity map after subtracting only the elastic peaks and Fig.~~\ref{fig:3}g shows the combined DMRG dynamical structure factors $S_{0}(q_{\parallel}, \omega)$, $S_{1}(q_{\parallel}, \omega)$, and $S_{2}(q_{\parallel}, \omega)$ obtained by fitting the RIXS spectra. The $\Delta S_0$ type of two-magnon continuum excitations provide a negligible contribution to the RIXS spectra (see Methods for the contributions from each type of excitation). Undoubtedly, the two-component excitation and its dispersion in the momentum space (Fig.~\ref{fig:3}e) is well described only by the quadrupolar $\Delta S_2$ excitation.

\section*{Quadrupolar excitations at finite $T$}
\begin{figure}[ht]
	\centering
	\includegraphics[width=1\linewidth]{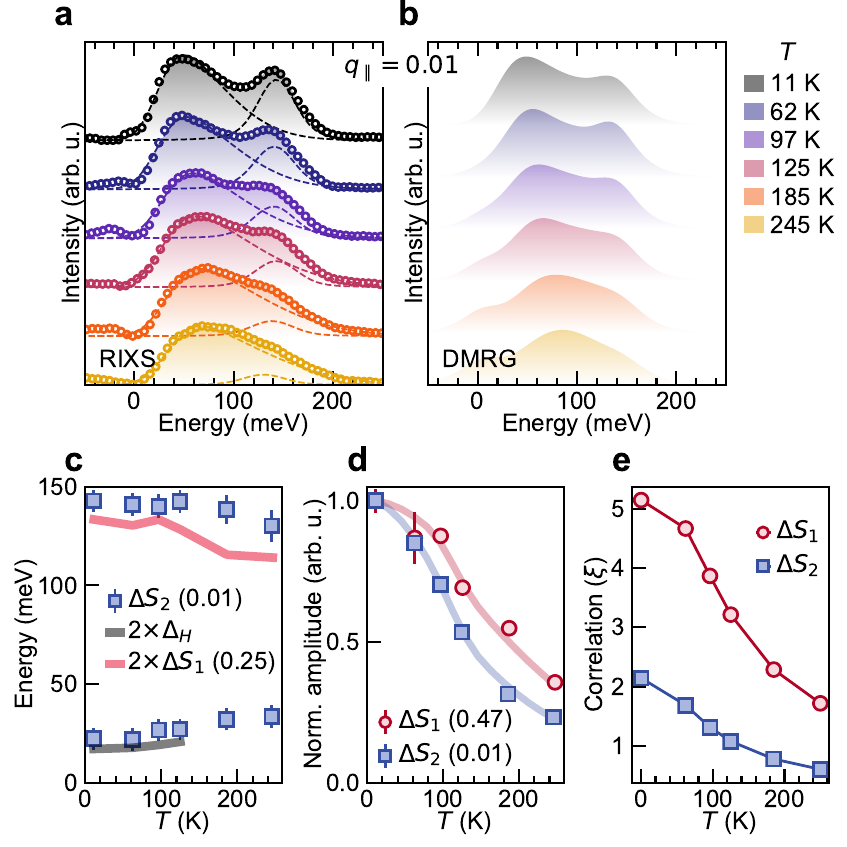}
	\caption{\textbf{Thermal stability and correlations of quadrupolar excitations in Y$_2$BaNiO$_5$.} \textbf{a,} RIXS spectra (open circles) at $q_{\parallel}$ = 0.01 with fitted elastic contributions subtracted, at different temperatures. The $\Delta S_2$ contributions have been fitted to extract the positions and amplitudes of the two branches shown by dashed lines (see Methods). The shaded regions show the complete fit profiles. \textbf{b,} $\Delta S_2$ profiles obtained from DMRG for corresponding temperatures. \textbf{c,} Energy variations with temperature of the two branches of the $\Delta S_2$ excitations extracted from RIXS shown in panel \textbf{a}. Black line shows the thermal variation of twice the Haldane gap energy obtained from Ref.~\cite{xu2007sci}. Red line shows the thermal variation of twice the $\Delta S_1$ peak energy at $q_{\parallel}$ = 0.25. \textbf{d,} Thermal variation in amplitudes of $\Delta S_1$ peak and the high energy peak of the $\Delta S_2$ extracted from fitted RIXS spectra at $q_{\parallel}$ = 0.47 and 0.01, respectively, normalised to corresponding amplitudes at 11~K. \textbf{e} Dipolar and quadrupolar correlation lengths for $\Delta S_\mathrm{tot}=1$ and 2 excitations obtained from DMRG static dipolar and quadrupolar correlation functions (see Methods).
	}
	\label{fig:4}
\end{figure}

To further understand the character of the two-component $\Delta S_2$ quadrupolar excitations, we also studied their thermal evolution. Fig.~\ref{fig:4}a shows the RIXS spectra at $q_{\parallel}=0.01$, with contributions from the $\Delta S_2$ excitations, for increasing temperatures. For comparison, Fig.~\ref{fig:4}b shows finite temperature DMRG simulations for the $S_{2}(q_{\parallel}=0.01, \omega)$ 
excitations, convoluted with the experimental energy resolution. Since  the raw DMRG data for the $\Delta S_2$ channel (Fig.~\ref{fig:2}f) contains a two-peak structure with an asymmetric peak appearing at low energy and a symmetric peak at high energy, we fit the experimental $S_{2}(q_{\parallel}=0.01, \omega)$ finite temperature data using two components (see methods).   
Our results show that only the $\Delta S_2$ excitations obtained from DMRG are needed to reproduce the experimental RIXS spectra, even at finite temperatures. As shown in Fig.~\ref{fig:4}c, the peak energy of the lower component increases with temperatures following twice of the system's Haldane gap from Ref.~\cite{xu2007sci}. Conversely, the high-energy peak begins to soften above the Haldane gap temperature of $~\sim 100$~K. The highest energy value of the $\Delta S_1$ peak at $q_{\parallel}=0.25$ also follows this trend. The bandwidth reduction of the $\Delta S_1$ triplet dispersion with temperature occurs due to the thermal blocking of propagation lengths and decoherence~\cite{kenzelmann2002prb2}. In a simple picture, if one considers a continuum from pairs of non-interacting triplets due to single spin-flips at multiple sites, then bandwidth reduction of each would manifest as the overall raising and lowering of the lower and upper boundaries of the continuum, respectively (see Supplementary Note 3). A similar thermal effect on the propagation of the quadrupolar $\Delta S_2$ excitation should occur. The spectral weight of the two components in the $\Delta S_2$ excitation, however, behave differently with temperature. The low-energy continuum intensity varies little, while the high-energy peak diminishes rapidly with increasing temperature and the rate of decay is comparable to the $\Delta S_1$ peak at $q_{\parallel}=0.47$ (see Fig.~\ref{fig:4}d). The DMRG calculated correlation lengths (in lattice units) of the $\Delta S_1$ and $\Delta S_2$ excitations, as shown in Fig.~\ref{fig:4}e, also decrease in a similar way with temperature. 

\section*{Dual nature of the quadrupolar excitations}
The energy-momentum and temperature dependence of the quadrupolar $\Delta S_2$ excitations implies that their low- and high-energy components are different in nature. At low-energy, a conceivable picture is that immediately after a pair of triplets are created by a single-site excitation, they decay into two non-interacting triplets (see Fig.~\ref{fig:1}d) propagating incoherently along the chain and giving rise to the broad low-energy continuum 
%resembling a two-magnon continuum with an energy gap of $\sim 2\Delta_H$ 
(see Fig.~\ref{fig:2}c for the expected continuum boundaries). This behaviour is remarkably similar to the fractionalisation of a $\Delta S_\mathrm{tot}=1$ excitation into a two-spinon continuum in isotropic spin-1/2 chains~\cite{mourigal2013np}, but having a distinct origin. 

The sharpness of the high-energy component and its rapid decay with temperature, on the other hand, hints that it behaves as a distinct quasiparticle formed from pairs of triplets propagating coherently (see Fig.~\ref{fig:1}d)~\cite{knetter2001prl,kohno2007np}. In low-dimensional systems, sharp peaks in the magnetic spectrum may either originate from a van Hove singularity in the density of states of quasiparticles (in our case, non-interacting triplet pairs) or from the formation of a bound state 
%of a pair of quasiparticles 
(in our case, bound triplet pairs). In Fig.~\ref{fig:2}c, the lower and upper boundaries of the continuum from pairs of non-interacting triplets (equivalent to the two-magnon continuum) are shown. The high-energy component of $\Delta S_2$ appears above the upper boundary of the continuum, ruling out the van Hove singularity scenario and suggesting the formation of a bound state. The peak energy of the high-energy component is slightly larger than twice the highest energy value of the $\Delta S_1$ peak (by $\sim 7$~meV at $T=11$~K) and, surprisingly, remains so up to the highest measured temperature. The small positive energy difference suggests a weak \emph{repulsive} interaction between the bound triplets formed after a quadrupolar $\Delta S_2$ excitation~\cite{white1993prb,tsao1975prb}. Supplementary Note 6 provides a semi-quantitative energy scale based argument to support the notion of the bound state of weakly repulsing triplets excitations.

The correlation length of the $\Delta S_1$ excitations in Haldane spin chains decay exponentially in the presence of a  non-local order, which can be viewed to originate from alternating $S_z=\pm1$ sites intervened by $S_z=0$ sites. The $\Delta S_2$ excitations also have an exponentially decaying correlation length, albeit smaller than $\Delta S_1$ excitations, most likely because the quadrupolar excitations can only occur at the $S_z=\pm1$ sites and not the $S_z=0$ sites. As such, they may provide a means to detect the hidden non-local order that, at present, is only estimated theoretically by considering the $S_z=\pm1$ states in Haldane spin chains~\cite{garcia2004prl,hilker2017sci,kennedy1992cmp}. Overall, we show that the $\Delta S_2$ excitations sustain the isotropic properties, the exchange interactions, and the coherence inherent to the system. 
% Astonishingly however, we also find that the $\Delta S_2$ excitations posses a dual nature of propagation arising from the same process. 

\section*{Outlook}
Magnetic excitations provide vital information about a system's thermodynamic, magneto-transport, ultrafast magnetic, spintronic, or superconducting properties. Moreover, higher-order multipolar spin and orbital degrees of freedom give rise to exotic non-classical phenomena like the Kitaev spin liquid~\cite{takagi2019nrp}, multi-spinons~\cite{Schlappa2018}, spin-nematicity~\cite{kohama2019pnas}
%~\cite{kohama2019pnas,ramos2018prb}
, on-site multiferroicity~\cite{akaki2017prb}, and bound-magnon states
%~\cite{bai2021np,dally2020prl,smerald2013prb}
~\cite{bai2021np, dally2020prl}. However, as noted earlier, multipolar excitations are challenging to detect using conventional probes~\cite{savary2015arx}.  It was shown recently that quadrupolar excitations in $S=1$ FeI$_2$ appear in INS only due to their hybridisation with dipolar excitations through anisotropic spin-exchanges~\cite{bai2021np}. In the presence of a strong anisotropy, bound magnetic excitations from $\Delta S_\mathrm{tot}=2$ spin-flips at energies higher than the two-magnon continuum have also been observed in $S=1$ spin chains using high-field electron spin resonance ~\cite{zvyagin2007prl,papanicolaou1997prb}. However, our demonstration of pure quadrupolar spin dynamics in an isotropic Haldane system without invoking anistropic interactions suggests that simultaneous confinement and propagation of excitations can occur entirely via higher-order quantum correlations~\cite{lake2010np}. Our work thus illustrates RIXS's capability of detecting higher-order propagating excitations, irrespective of the presence of dipolar excitations~\cite{bai2021np}. This method may be the only way to study quadrupolar excitations in spin-nematic systems where the dipolar excitations are suppressed~\cite{smerald2013prb}. On the other hand, exploring the physics of high-energy excitations and/or eigenstates of a simple \emph{non-integrable} spin-chain model is also of great theoretical interest. We provide a simple physical intuition about the nature of one of the high-energy eigenstates of the one-dimensional Heisenberg model and our findings may have important consequences for many-body localized states of matter and thermalisation in quantum systems\cite{nandkishore2015many}. Looking forward, it would be interesting to learn how the quadrupolar excitations can be manipulated with intrinsic perturbations like anisotropy or extrinsic ones like a magnetic field. The indications of a propagating quadrupolar \textit{bound} excitation in a real material can also have important ramifications for realising quantum information transfer in form of qubit pairs~\cite{subrahmanyam2004pra,bose2013nat}.
%~\cite{ghosh2014pra,subrahmanyam2004pra,bose2013nat}. 
% Belonging to the same universality class of the non-integrable Affleck-Kennedy-Lieb-Tasaki (AKLT) model\cite{affleck1988cmp}, where recent efforts have shown that a closed-form expression for selected excited states can be provided\cite{Moudgalya2018}.

\section*{Methods}
\subsection*{Experiments}
A single crystal of Y$_2$BaNiO$_5$ grown by the floating-zone method was used for the RIXS measurements. The momentum transfer along the chain direction $q_{\parallel}$ was varied by changing the x-ray incident angle $\theta$ while keeping the scattering angle fixed at 154$^{\circ}$. The lattice constant along the chain or the $c$-axis used for calculation of momemtum transfer is 3.77 \AA. The crystal was cleaved in vacuum and the pressure in the experimental chamber was maintained below $\sim5\times10^{-10}$ mbar throughout the experiment. High energy-resolution RIXS data ($\Delta E\simeq$ 37~meV) at the Ni $L_3$-edge were collected at I21 RIXS beamline, Diamond Light Source, United Kingdom~\cite{i21web}. The zero-energy position and resolution of the RIXS spectra were determined from subsequent measurements of elastic peaks from an adjacent carbon tape. The polarization vector of the incident x-ray was parallel to the scattering plane (i.e. $\pi$ polarization). See Supplementary Note 1 for more details of the experimental configuration.  

\subsection*{Theory}
In the main text, we pointed out that the $S=1$ Haldane chain system Y$_2$BaNiO$_5$ might present negligibly small single-ion anisotropy $\sim0.035J$, exchange-anisotropy $\sim0.011J$ terms in a Heisenberg model description at low energies. We have verified numerically that these small corrections do not change our magnetic spectra qualitatively, and therefore a pure isotropic Heisenberg model has been adopted throughout our study.
\subsubsection*{Zero and finite temperature DMRG Calculations}
$T=0$ DMRG calculations on 100 site chains with open boundary conditions (OBC) were carried out with the correction-vector method\cite{PhysRevB.60.335} using the Krylov decomposition\cite{PhysRevE.94.053308}, as implemented in the DMRG++ code\cite{alvarez0209}. This approach requires real-space representation for the dynamical structure factors in the frequency domain, which can be found in the Supplementary Note 4. %\subsubsection*{Finite temperature DMRG Calculations}
For $T>0$ calculations we used the ancilla (or purification) method with a system of 32 physical and 32 ancilla sites, also with OBC. For more details see Supplementary Note 7.
%:  We first prepare the system in a $T \rightarrow \infty$ (inverse temperature $\beta=1/T= 0$) state using a fictitious ``entangler'' Hamiltonian $H_E$ acting in an enlarged Hilbert space. In the purification approach, the enlarged system (physical and ancilla sites) can be described as a spin ladder, with physical sites on one leg and ancilla sites on the other. The entangler Hamiltonian is chosen such that its ground state $|\psi_\infty \rangle$ corresponds to the $T\rightarrow \infty$ state  of the physical system when the ancilla degrees of freedom are traced out, $\rho={\rm Tr}\left[|\psi_\infty\rangle\langle\psi_\infty|\right]_A$.
%Second, the system is cooled through evolving in imaginary time with the physical Hamiltonian $H$ acting only on physical sites $e^{-\beta(H\otimes I)/2}$, where $I$ is the identity operator in the ancilla space. To compute the thermal state of the system $|\psi(\beta)\rangle={e^{-\beta(H\otimes I)/2}}|\psi_\infty\rangle$, we used the Krylov algorithm for time evolution with an imaginary time step $\Delta\beta=0.025/J$.
%Finally, the dynamical spin structure factor is calculated using the operator ${\cal L}=H\otimes I+I\otimes(-H)$ with the real space expression reported in the Supplementary materials.

\noindent
For both the zero temperature and finite temperature calculations we kept up to $m=2000$ DMRG states to maintain a truncation error below $10^{-7}$ and $10^{-6}$, respectively and introduced a spectral broadening in the  correction-vector approach fixed at $\eta=0.25J=6$~meV.

\subsubsection*{Dynamical spin correlation functions}
We consider three correlation functions $S_0(q_{\parallel},\omega)$, $S_1(q_{\parallel},\omega)$, and $S_2(q_{\parallel},\omega)$ giving information about $\Delta S_\mathrm{tot}=0$, $\Delta S_\mathrm{tot}=1$, and $\Delta S_\mathrm{tot}=2$ excitations, respectively.
To make the expressions more transparent, we use the Lehmann representation and construct the corresponding excitation operators in momentum space. The relevant correlation functions are 
\begin{equation}
    S_{\alpha}(q_{\parallel},\omega)=\sum_f |\langle f|S^{\alpha}_{q_{\parallel}}|\psi\rangle|^2 \delta(\omega-E_f +E_{\psi}),\label{eqn:corr}
\end{equation}
where $|f\rangle$ are the final states of the RIXS process and
\begin{align}
S^{0}_{q_{\parallel}}&=\frac{1}{\sqrt{L}}\sum_{j} e^{i q_{\parallel} j} \vec{S}_j\cdot\vec{S}_{j+1}\label{eqn:ops0},\\
S^{1}_{q_{\parallel}}&=\frac{1}{\sqrt{L}}\sum_{j} e^{i q_{\parallel} j} S^{+}_j\label{eqn:ops1},\\
S^{2}_{q_{\parallel}}&=\frac{1}{\sqrt{L}}\sum_{j} e^{i q_{\parallel} j} (S^{+}_j)^2.\label{eqn:ops2}
\end{align}
\noindent
The three dynamical correlation functions given by Eq.~(\ref{eqn:corr}) appear at the lowest order of a ultrashort core-hole lifetime expansion of the full RIXS cross-section. As the single-site $\Delta S_\mathrm{tot}=0$ RIXS scattering operator is trivial (identity operator) 
in a low-energy description in terms of spin $S=1$ sites,
the lowest order operator would involve two-sites and, 
by rotational symmetry, involves a scalar product of neighbouring spin operators [see Eq.~(\ref{eqn:ops0})].
Single and double spin-flip RIXS scattering operators, on the other hand, lead to $\Delta S_\mathrm{tot}=1$ and $\Delta S_\mathrm{tot}=2$ excitations and can be naturally described in terms of onsite $S^{+}_j$ and $(S^{+}_j)^2$ operators, respectively [Eqs.~(\ref{eqn:ops1}) \& (\ref{eqn:ops2})]. In the Supplementary Note 5 we provide analysis of the three dynamical correlation functions in terms of single triplet excitations or \emph{magnon} states in the Haldane chain.

\subsubsection*{Correlation lengths}
Figure~\ref{fig:4}e of the main text shows dipolar and quadrupolar correlation lengths as a function of temperature. These have been obtained by computing $\langle\psi(\beta)| S^{+}_j S^{-}_{j+r}|\psi(\beta)\rangle$ and $\langle \psi(\beta)|(S^{+}_j)^2(S^{-}_{j+r})^2|\psi(\beta)\rangle$ correlation functions from the center of the chain $j=c$, respectively, and fitting with a exponential decay relationship $f(r)=A e^{-r/\xi}$.

\subsection*{RIXS data fitting}
RIXS data were normalised to the incident photon flux and corrected for x-ray self-absorption effects prior to fitting. The elastic peak was fit with a Gaussian function with a width set by the energy resolution. The RIXS spin excitations in Fig.~\ref{fig:3} were modeled with the Bose factor weighted dynamical spin susceptibilities obtained from our $T=0$ DMRG calculations for $\Delta S_\mathrm{tot}=0$, 1, and 2 ($S_0$, $S_1$ and $S_2$, respectively) after they were convoluted with a Gaussian function capturing the experimental energy resolution.
The total model intensity is given by
\begin{equation}
\begin{aligned}
	I_{\mathrm{RIXS}}(q_{\parallel}, \omega) = C_0(q_{\parallel})S_0(q_{\parallel}, \omega)\\+C_1(q_{\parallel})S_1(q_{\parallel}, \omega)+C_2(q_{\parallel})S_2(q_{\parallel}, \omega),
\label{eqn:irixs}
\end{aligned}
\end{equation}
\noindent
where the coefficients $C_0$, $C_1$ and $C_2$ account for the varying RIXS scattering cross section for each spin excitation with varying $\theta$ ($q_{\parallel}$). The reader is referred to the Supplementary Note 2 for the extracted values of coefficients and the fit profiles. As seen in Fig.~\ref{fig:3}c, $S_2$ from DMRG for $\Delta S_\mathrm{tot}=2$ transitions do not capture the additional spectral weight on the high energy side in the RIXS signal. A `half'-Lorentzian truncated damped harmonic oscillator (HLDHO) function centered at 0.175 eV was therefore included in the fits to account for this tail. In Fig.~\ref{fig:4}, the $\Delta S_\mathrm{tot}=2$ excitations at $q_{\parallel}=0.01$ $(2\pi/c)$ are decomposed using a DHO function for the sharp high energy peak and a skewed Gaussian function for the broad lower continuum. Both functions are energy-resolution convoluted and weighted by a Bose factor. DHO functions were used similarly for fitting $\Delta S_\mathrm{tot}=1$ triplet excitations at $q_{\parallel}=0.47$ (see Supplementary Note 2).

\section*{Data availability}
%The data supporting this study can be found at \url{Insert Url.}
The data supporting this study can be made available upon request.

\section*{Code availability}
The numerical results reported in this work were obtained with DMRG++ versions 6.01 and PsimagLite versions 3.01. The DMRG++ computer program~\cite{alvarez0209} is available at \url{https://github.com/g1257/dmrgpp.git} (see Supplementary Note 7 for more details.)

\section*{Acknowledgements}
We thank I. Affleck, J.-G. Park, S. Hayden, A. Aligia, and K. Wohlfeld for insightful discussions. We are grateful to C. Bastista for suggesting the possibility of a van Hove singularity in our results. All data were taken at the I21 RIXS beam line of Diamond Light Source (United Kingdom) using the RIXS spectrometer designed, built, and owned by Diamond Light Source. We acknowledge Diamond Light Source for providing the beam time on beam line I21 under Proposal MM24593. S. J. acknowledges support from the National Science Foundation under Grant No. DMR-1842056. A. Nocera acknowledges support from the Max Planck-UBC-UTokyo Center for Quantum Materials and Canada First Research  Excellence Fund (CFREF) Quantum Materials and Future Technologies Program of the Stewart Blusson Quantum Matter Institute (SBQMI), and the Natural Sciences and Engineering Research Council of Canada (NSERC). This work used computational resources and services provided by Compute Canada and Advanced Research Computing at the University of British Columbia. We acknowledge T. Rice for the technical support throughout the beam times. We also thank G. B. G. Stenning and D.W. Nye for help on the Laue instrument in the Materials Characterisation Laboratory at the ISIS Neutron and Muon Source.

\section*{Author contributions}
K.-J.Z. conceived the project; K.-J.Z., A. Nag, A. Nocera, and S.J. supervised the project. A. Nag, K.-J.Z., S.A., M.G.-F., and A.C.W. performed RIXS measurements. A. Nag, S.A., and K.-J.Z. analysed RIXS data. S.-W.C. synthesized and characterised the sample. A. Nocera and S.J. performed DMRG calculations. A. Nag, K.-J.Z., A. Nocera, and S.J. wrote the manuscript with comments from all the authors.

\section*{Competing interests}
The authors declare no competing interests. 

\section*{Additional information}
\textbf{Supplementary Information} is available for this paper. Correspondence and requests for materials should be addressed to A. Nag or A. Nocera or S.J. or K.-J.Z.

\bibliography{references2}

\end{document}

\title{Supplementary Information for quadrupolar magnetic excitations in an isotropic spin-1 antiferromagnet}

\author{A. Nag}\email[]{abhishek.nag@diamond.ac.uk}
\affiliation{Diamond Light Source, Harwell Campus, Didcot OX11 0DE, United Kingdom}

\author{A. Nocera}\email[]{alberto.nocera@ubc.ca}
\affiliation{Stewart Blusson Quantum Matter Institute, University of British Columbia, Vancouver, British Columbia, V6T 1Z4 Canada}
\affiliation{Department of Physics Astronomy, University of British Columbia, Vancouver, British Columbia, Canada V6T 1Z1}

\author{S. Agrestini}
\author{M. Garcia-Fernandez}
\author{A. C. Walters}
\affiliation{Diamond Light Source, Harwell Campus, Didcot OX11 0DE, United Kingdom}

\author{Sang-Wook Cheong}
\affiliation{Rutgers Center for Emergent Materials, Rutgers University, Piscataway, NJ, USA.}

\author{S. Johnston}\email[]{sjohn145@utk.edu}
\affiliation{Department of Physics and Astronomy, The University of Tennessee, Knoxville, Tennessee 37966, USA}

\author{Ke-Jin Zhou}\email[]{kejin.zhou@diamond.ac.uk}
\affiliation{Diamond Light Source, Harwell Campus, Didcot OX11 0DE, United Kingdom}

\renewcommand{\figurename}{{\bf Supplementary Figure}}
\renewcommand{\thefigure}{{S\arabic{figure}}}%
\renewcommand{\theequation}{S\arabic{equation}}%
\maketitle

% \begin{center}
%     \Large{\bf Supplemental Information}
% \end{center}
\section*{Supplementary Note 1: Experimental details}
\begin{figure}[ht]
\centering
\includegraphics[width=1\linewidth]{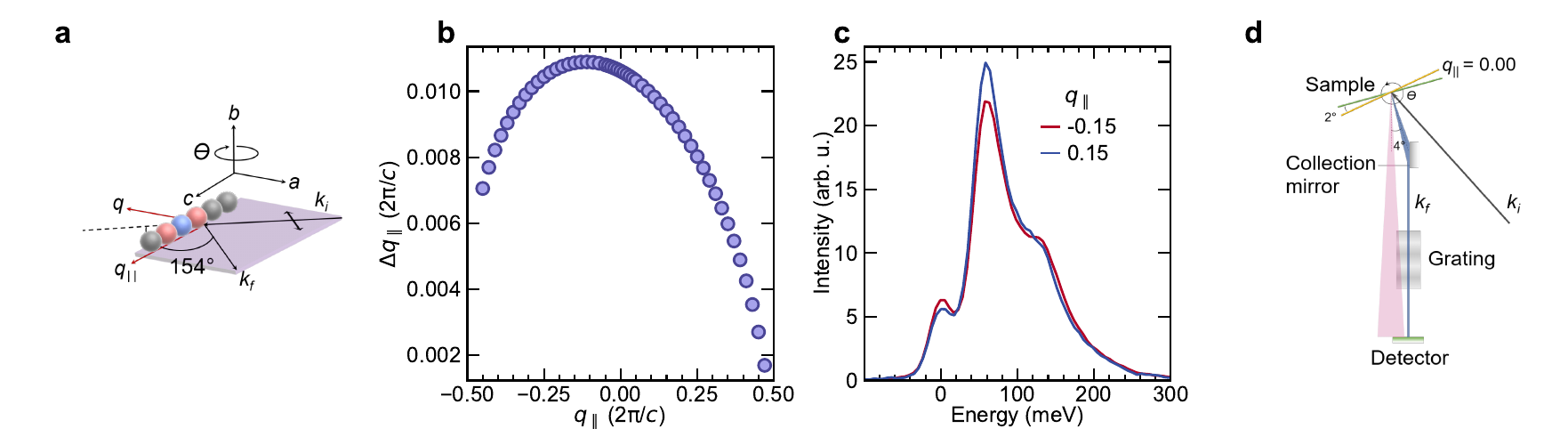}
\caption{\textbf{a} RIXS geometry for probing spin excitations in Y$_2$BaNiO$_5$. Throughout the paper, projection of  momentum transfer ($q_{\parallel}$) in units of $2\pi/c$ along the chain direction is presented. \textbf{b} The momentum transfer resolution for our experimental setup as a function of $q_\parallel$. \textbf{c} A comparison of the RIXS spectra recorded at $\pm q_\parallel$. The agreement between the two RIXS spectra confirm that our assigned $q_{\parallel}=0.00$ corresponds to the true specular reflection condition. \textbf{d} A schematic representation of the optical arrangement for scattered X-rays from sample at I21 RIXS beamline, Diamond Light Source. The $k_f$ and true momentum transfer values are determined by the scattered beam collected by a plane parabolic mirror $\sim 4^{\circ}$ higher than the direct scattered beam direction. The incidence angle on the sample for specular reflection condition for the direct beam is, therefore, $\sim 2^{\circ}$ lower than that for the true specular reflection condition at $q_{\parallel}=0.00$. Usually this direct beam is weak and does not reach the detector. However, due to the high quality mirror like cleaved surface of the crystal, a strong specular elastic reflection corresponding to this direct beam is seen close to $q_{\parallel}=-0.03$ in Fig.~2(a) of the main text. 
}
\label{fig:EFig_qres}
\end{figure}

The single crystal of Y$_2$BaNiO$_5$ was pre-aligned using Laue diffraction so that its chain direction lies in the RIXS scattering plane (Fig.~\ref{fig:EFig_qres}a). The momentum transfer resolution is less than 0.011 ($2\pi/c$) (see Fig.~\ref{fig:EFig_qres}b). Negative and positive values of $q_{\parallel}$ represent  x-ray grazing-incident and  grazing-exit geometries, respectively. The strongest specular elastic reflection was seen close to $q_{\parallel}=-0.03$, as shown in Fig.~2(a) of main text. This position originates from a combination of the high quality mirror-like cleaved surface of the crystal and the arrangement of optical elements used for collecting the  scattered x-rays (see Fig.~\ref{fig:EFig_qres}c, d). 

\clearpage
\newpage
\section*{Supplementary Note 2: RIXS data fitting}
\begin{figure}[ht]
\centering
\includegraphics[width=1\linewidth]{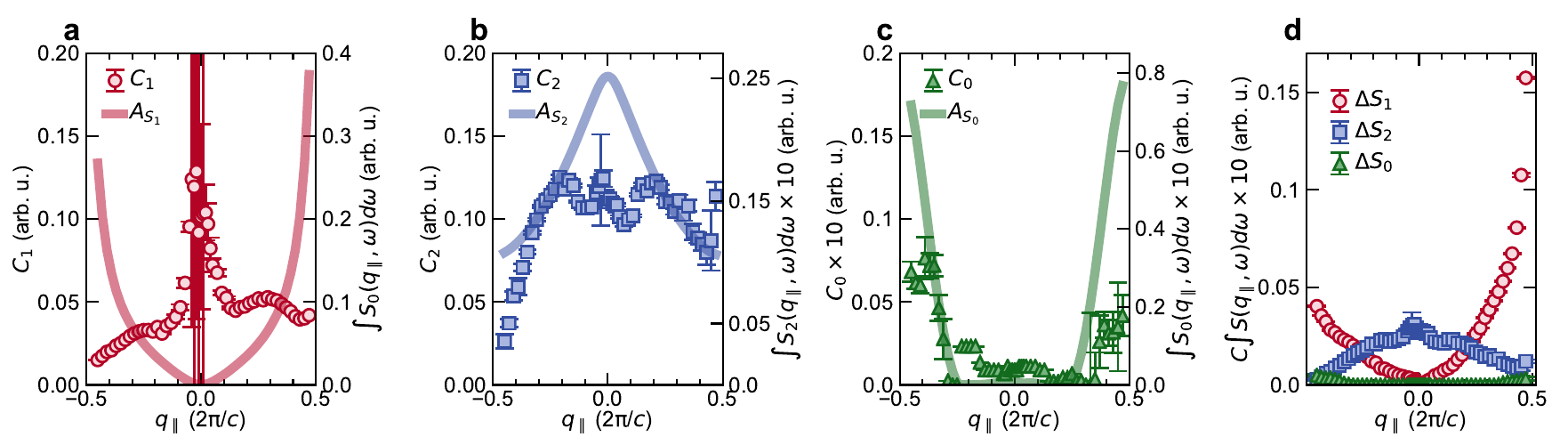}
\caption{\textbf{a, b, c} Coefficients $C_1$, $C_2$, and $C_0$ accounting for changes in the RIXS scattering cross section with varying x-ray incidence for $\Delta S_1$, $\Delta S_2$, and $\Delta S_0$ excitations, respectively. The coefficient values are determined by fitting RIXS spectra with dynamical spin susceptibilites obtained from DMRG using Eq.~(5) given in Methods. Error bars are least square fit errors. Also shown by the continuous lines are the  integrated dynamical spin structure factors $A_S(q_{\parallel})=\int S(q_{\parallel},\omega) d\omega$ from DMRG for the $\Delta S_1$, $\Delta S_2$, and $\Delta S_0$ excitations. \textbf{d} Total RIXS intensities for $\Delta S_1$, $\Delta S_2$, and $\Delta S_0$ excitations.
}
\label{fig:EFig_coeff}
\end{figure}

\begin{figure}[hb]
\centering
\includegraphics[width=1\linewidth]{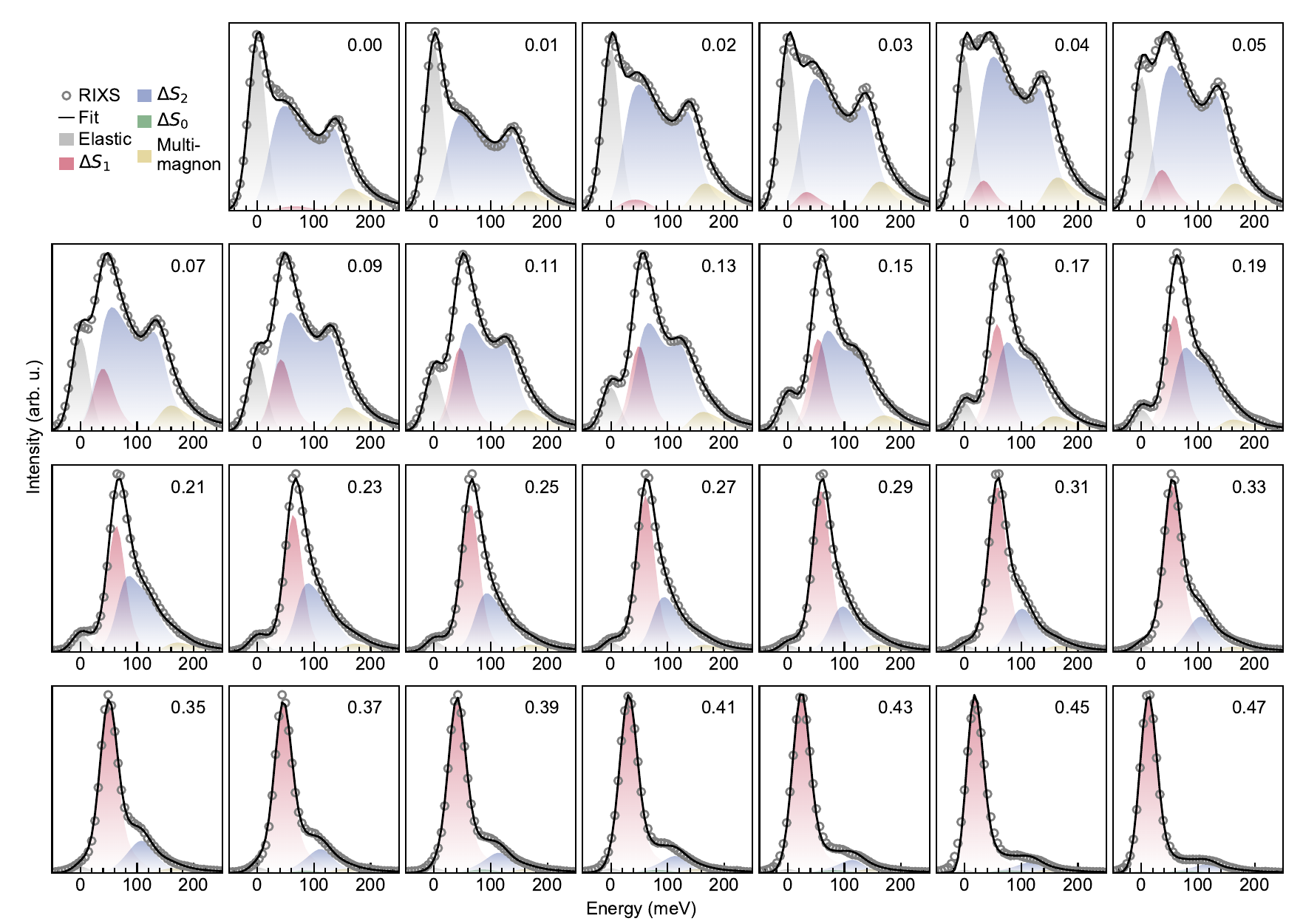}
\caption{Ni $L$-edge RIXS line spectra from Y$_2$BaNiO$_5$ at 11~K, with fits derived from the spin susceptibilities obtained from DMRG using Eq.~(5) given in Methods. Corresponding $q_{\parallel}$ value is written in each panel.  
}
\label{fig:EFig_qfits}
\end{figure}

\begin{figure}[ht]
\centering
\includegraphics[width=1\linewidth]{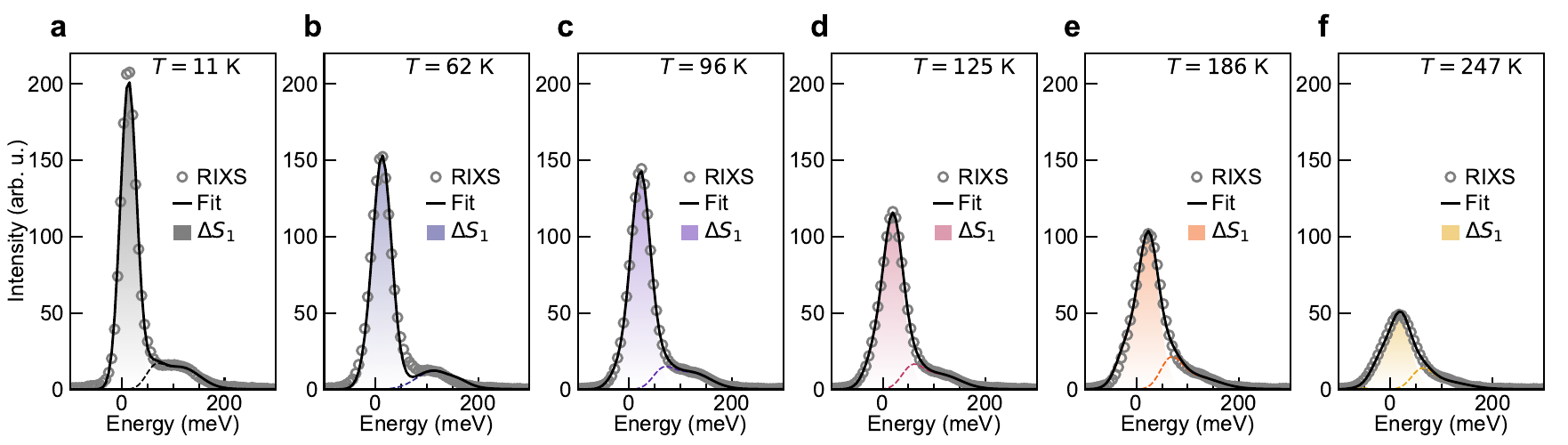}
\caption{\textbf{a-f} Temperature dependent RIXS spectra at $q_{\parallel}=0.47$. Shaded regions are fits to $\Delta S_\mathrm{tot}=1$ triplet excitations using resolution-convoluted damped harmonic oscillator functions weighted by the Bose factor. The extracted amplitudes have been used in Fig.~4d of main text. The dashed lines indicate contributions from the $\Delta S_\mathrm{tot}=2$ and multi-magnon excitations.
}
\label{fig:EFig_q0p47}
\end{figure}

Fig.~\ref{fig:EFig_coeff} shows the extracted values of coefficients $C_0$, $C_1$ and $C_2$ of the three different types of excitations considered for fitting the RIXS data as described in Methods of main text. Fig.~\ref{fig:EFig_qfits} shows the fit profiles of the different components. Fig.~\ref{fig:EFig_q0p47} shows the fitting of the $\Delta S_\mathrm{tot}=1$ triplet excitations at $q_{\parallel}=0.47$.

\clearpage
\newpage
\section*{Supplementary Note 3: Influence of temperature on the boundaries of two-$\Delta S_1$ excitations.}
\begin{figure}[ht]
\centering
\includegraphics[width=0.5\linewidth]{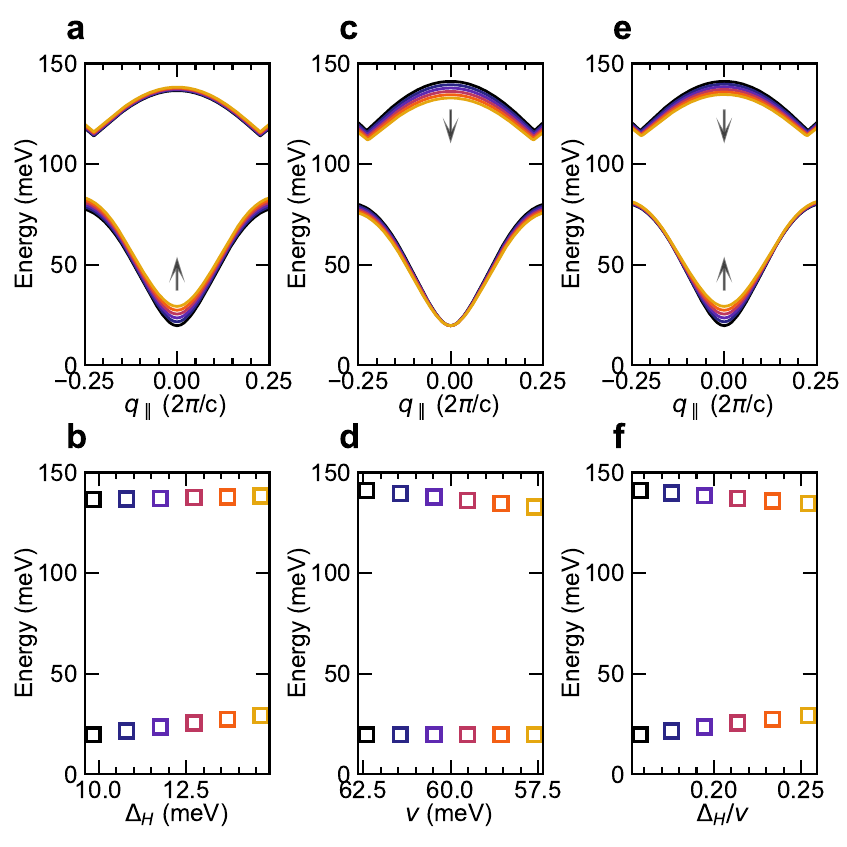}
\caption{\textbf{a}, The influence of Haldane gap $\Delta_H$ on the boundaries of two-$\Delta S_1$ excitations. The arrow indicates the hardening of the lower boundary with increasing $\Delta_H$. \textbf{b,} The variation in the energies of the upper and lower boundaries at $q_{\parallel}=0$ with increasing $\Delta_H$ extracted from the curves in panel \textbf{a}. \textbf{c} The influence of the velocity $v$ of the $\Delta S_1$ excitations on the boundaries of two-$\Delta S_1$ excitations. The arrow indicates the softening of the upper boundary with decreasing $v$. \textbf{d,} The variation in the energies of the upper and lower boundaries at $q_{\parallel}=0$ with decreasing $v$ extracted from the curves in panel \textbf{c}. \textbf{e} The influence of $\Delta_H$ and $v$ on the boundaries of the two $\Delta S_1$ excitations. The arrow indicates the hardening (softening) of the lower (upper) boundary with increasing (decreasing) $\Delta_H$ ($v$). \textbf{f,} The variation in the energies of the upper and lower boundaries at $q_{\parallel}=0$ with increasing (decreasing) $\Delta_H$ ($v$), extracted from the corresponding curves in panel \textbf{e}.  
}
\label{fig:EFig_2m_var}
\end{figure}

In a simple picture, if one considers a continuum from pairs of non-interacting $\Delta S_1$ triplets due to single spin-flips at multiple sites, then bandwidth reduction of each would manifest as the overall raising and lowering of the lower and upper boundaries of the continuum, respectively. In Fig.~\ref{fig:EFig_2m_var}a and b, we show the change of such continuum boundaries upon changing the Haldane gap energy $\Delta_H$ using the semi-quantitative Haldane dispersion relation given in the caption of Fig.~2 of main text for two $\Delta S_1$ triplets. In Fig.~\ref{fig:EFig_2m_var}c and d, we show the change of the continuum boundaries upon changing the velocity $v$ of the $\Delta S_1$ triplets and in Fig.~\ref{fig:EFig_2m_var}e and f, the combined effect of $\Delta_H$ and $v$. As temperature is raised, due to the increase (decrease) of the $\Delta_H$ ($v$), the energy of the lower (upper) boundary of the continuum at $q_{\parallel}=0$ increases (decreases). Similar effect is seen on the two components of the $\Delta S_2$ excitations in Fig.~4 of main text.

\clearpage
\newpage
\section*{Supplementary Note 4: Real-space representations for the dynamical structure factors}
Here, we provide the real-space representation of the dynamical structure factors studied in the main text. 
For a finite system, the dynamical structure factor is given by
\begin{equation}\label{eqn:dynfactor}
    S_{\alpha}(q_{\parallel},\omega)=\frac{1}{L}\sum^{L}_{j=1}e^{\mathrm{i}q_{\parallel}(j-c)}\Big\{-\frac{1}{\pi}{\rm Im}\Big[\langle\psi|S^{\alpha}_{j}\frac{1}{\omega-H+\mathrm{i}\eta}S^{\alpha}_{c}|\psi\rangle\Big]\Big\},
\end{equation}
where the operator $S^{\alpha}_{j}$ is applied at the center $j=c$ of the long chains with open boundaries\cite{White2004,PhysRevE.94.053308,Nocera2018b}, and $|\psi\rangle$ represents the ground state of the system.

Using the Liouville\cite{Tiegel2013} and the purification approaches\cite{Feiguin2005,Feiguin2010,Nocera2016}, 
the dynamical structure factor in real space and a finite temperature $T=1/\beta$ is
\begin{equation}\label{eq:Liouville}
    S_{\alpha}(q_{\parallel},\omega)=\frac{1}{L}\sum^{L}_{j=1}e^{\mathrm{i}q_{\parallel}(j-c)}\Big\{-\frac{1}{\pi}{\rm Im}\Big[\langle\psi(\beta)|S^{\alpha}_{j}\frac{1}{\omega-{\cal L}+\mathrm{i}\eta}S^{\alpha}_{c}|\psi(\beta)\rangle\Big]\Big\},
\end{equation}
where ${\cal L}=H\otimes I+I\otimes(-H)$ is the Liouville operator and 
$|\psi(\beta)\rangle$ is the purified thermal state given by $|\psi(\beta)\rangle={e^{-\beta(H\otimes I)/2}}|\psi_\infty\rangle$. Here, the Hamiltonian acts only on the physical sites of the system and $|\psi_\infty\rangle$ is the maximally entangled infinite temperature state (more details are provided in the Supplementary Note 4).

\section*{Supplementary Note 5: Magnon-like states unveiled by the dynamical spin correlations in Haldane spin chains}

In this section we analyze the excitations encoded in the three dynamical correlation functions introduced in the main text in Eq.~(1) in terms of single triplet excitations or \emph{magnon} states in the Haldane chain. Following Eq.~(5) in Ref.~\cite{white1993prb}, a single magnon state in an infinite chain with momentum $q_{\parallel}$ and spin quantum number $S^{z}=\alpha$ has the form 
\begin{equation}
    M_{q_{\parallel},\alpha}|\psi\rangle=|q_{\parallel},\alpha\rangle = \sum_{j} e^{\mathrm{i} q_{\parallel} j}c^{\dag}_j(q_{\parallel},\alpha)|\psi\rangle,
\end{equation}
 where the operator $c^{\dag}_j(q_{\parallel},\alpha)$ includes the $S^{+}_j$
operator and products of multiple spin operators in the vicinity of site $j$. In other words, single magnons are not constructed by applying only local spin operators $S^{1}_{q_{\parallel}}=\frac{1}{\sqrt{L}}\sum_{j} e^{\mathrm{i} q_{\parallel} j} S^{+}_j$. When $S^{1}_{q_{\parallel}}$ is applied to the ground state, the resulting state can be expressed at low energy by a multi-magnon expansion 
\begin{equation}
    S^{1}_{q_{\parallel}}|\psi\rangle= \Big(\sum_{q_{\parallel},\alpha} t^{1}(q_{\parallel},\alpha)M_{q_{\parallel},\alpha}+\sum_{q^{1}_{\parallel},\beta,q^{(2)}_{\parallel},\gamma} s^{1}(q^{(1)}_{\parallel},\beta,q^{(2)}_{\parallel},\gamma)M_{q^{(1)}_{\parallel},\beta}M_{q^{(2)}_{\parallel},\gamma} + ...\Big)|\psi\rangle,
\end{equation}
where two-magnon and higher-order multi-magnon states naturally appear. Note that $t^{1}(q_{\parallel},\alpha)$ and $s^{1}(q^{(1)}_{\parallel},\beta,q^{(2)}_{\parallel},\gamma)$ encapsulate the projection of the $S^{1}_{q_{\parallel}}|\psi\rangle$ state on the single- and two-magnon wave functions, respectively. 
The expression above tells us that the dynamical correlation function $S_{1}(q_{\parallel},\omega)$ encodes information about single magnon excitations, but also two-magnon and three-magnon excitations as well~\cite{Affleck1999,Essler2000} (with a reduced spectral weight). By symmetry arguments, it is easy to prove that $S^{0}_{q_{\parallel}}$ and $S^{2}_{q_{\parallel}}$ operators can only induce states with single or multiple pairs of magnon excitations
\begin{align}
    S^{0}_{q_{\parallel}}|\psi\rangle&=  \Big(t^{0}(q_{\parallel})+\sum_{q^{1}_{\parallel},\beta,q^{(2)}_{\parallel},\gamma} s^{0}(q^{(1)}_{\parallel},\beta,q^{(2)}_{\parallel},\gamma)M_{q^{(1)}_{\parallel},\beta}M_{q^{(2)}_{\parallel},\gamma} + ...\Big)|\psi\rangle\\
S^{2}_{q_{\parallel}}|\psi\rangle&= \Big(\sum_{q^{1}_{\parallel},\beta,q^{(2)}_{\parallel},\gamma} s^{2}(q^{(1)}_{\parallel},\beta,q^{(2)}_{\parallel},\gamma)M_{q^{(1)}_{\parallel},\beta}M_{q^{(2)}_{\parallel},\gamma} + ...\Big)|\psi\rangle.   
\end{align}
Here, $s^{0(2)}(q^{(1)}_{\parallel},\beta,q^{(2)}_{\parallel},\gamma)$ describes the projection of the $S^{0(2)}_{q_{\parallel}}|\psi\rangle$ states on the two-magnon wave function. 

Note that the dynamical correlator $S_{0}(q_{\parallel},\omega)$ defined in the main text naturally contains an elastic peak since the $S^{0}_{q_{\parallel}}$ operator is equal to the Hamiltonian operator at $q_{\parallel}=0$ and $t^{0}(q_{\parallel} =0)=E_{\rm gs}$.

\section*{Supplementary Note 6: Bound state of $\Delta S_2$ excitations}

In this section, we discuss the physical origin of the sharp peak observed in the  $\Delta S_2$ spectrum close to $q_{\parallel}=0$ with an energy of $\simeq136$~meV (see Figs.~2e and 2f of the main text). In low-dimensional systems, a van Hove singularity in  magnetic density of states can give rise to sharp peaks at momentum values where the magnetic dispersion has  stationary points. The sharp peaks observed in our study could therefore originate either from a van Hove singularity in the triplet  density of states, or from the formation of a bound state of the triplet pairs \emph{above} the upper boundary of the continuum formed (equivalent to two-magnons). 
\begin{figure}[ht]
	\centering
	\includegraphics[width=0.5\linewidth]{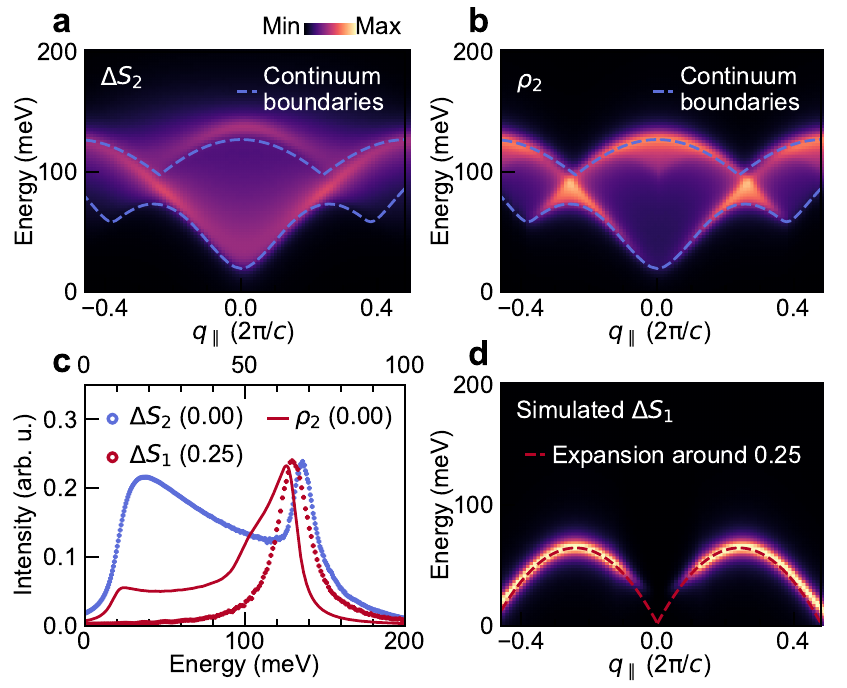}
	\caption{\textbf{a,} $S_2(q_{\parallel},\omega)$ from DMRG. \textbf{b,} $\rho_{2}(q_{\parallel},\omega)$ obtained from joint $\Delta S_1$ density of states. Both panels are over plotted with the lower and upper boundaries of the continuum from pairs of non-interacting triplets (equivalent to the two-magnon continuum). \textbf{c,}  $S_2(q_{\parallel}=0,\omega)$, $\rho_2(q_{\parallel}=0,\omega)$ and $S_1(q_{\parallel}=0.25,2\omega)$. The top $x$-axis is the energy scale for $S_1(q_{\parallel}=0.25,\omega)$. \textbf{d,} One magnon dispersion relationship fitted from DMRG and experimental RIXS data. We expand it around $q_{\parallel}\simeq\pm 0.25$ obtaining free quasiparticles with \emph{negative} mass. Here negative mass just means that the triplet energy cannot be bigger than $\omega_{S_1}(q_{\parallel}\simeq0.25)=64.5$~meV.
	}
	\label{fig:EFig_vH}
\end{figure}

Adopting the dispersion of $\Delta S_1$ triplet excitations~\cite{xu1996prb,zaliznyak2001prl,xu2007sci,Santos1989}  obtained from our DMRG calculations  $\omega^2_{S_1}(q_{\parallel})=\Delta^2_\mathrm{H}+v^2\mathrm{sin}^2q_{\parallel}+\alpha^2\mathrm{cos}^2\frac{q_{\parallel}}{2}$, stationary points are located at 
$q_{\parallel}=0$ and $\simeq0.25$ ($2\pi/c$) (see Fig.~\ref{fig:EFig_vH}d). At $q_{\parallel}\simeq0.25$, where the $\Delta S_1$ excitation has the highest energy and thus can contribute to the highest triplet pair energy, the $\Delta S_1$ density of states has a square-root singularity $\sim (|\omega-\omega_{S_1}(q_{\parallel}\simeq0.25)|)^{-1/2}$. Ignoring matrix elements effects, the  triplet pair density of states can be assumed to be proportional to the  \emph{joint} $\Delta S_1$ density of states
$\rho_2(q_{\parallel},\omega)=\sum_{k_{\parallel}}\delta(\omega-\omega_{S_1}(k_{\parallel})-\omega_{S_1}(q_{\parallel}-k_{\parallel}))$, which is shown in Fig.~\ref{fig:EFig_vH}b. Fig.~\ref{fig:EFig_vH}c shows the DMRG calculated $S_2(q_{\parallel}=0,\omega)$ with the sharp peak at an energy of $\sim136$~meV. Compared to it, $\rho_2(q_{\parallel}=0,\omega)$ has an asymmetric peak corresponding to the van Hove singularity at $\sim125$~eV (lower by $\sim11$ meV). Evidently, the sharp component of the $\Delta S_2$ excitation does not arise from the singularities in the magnetic density of states if the triplets are non-interacting and therefore have negligible bandwidth renormalisation.

In Fig.~\ref{fig:EFig_vH}c, $S_1(q_{\parallel}=0.25,2\omega)$ is plotted by doubling the energy scale, to show that the maximum energy of a pair of non-interacting triplets can be $~129$~meV ($\sim64.5\times2$, the true energy scale for $S_1(q_{\parallel}=0.25,\omega)$ is given on top of the panel). The sharp peak obtained in the $\Delta S_2$ excitations is therefore $\sim7$~meV higher than the non-interacting scenario at $q_{\parallel}=0$. Also, shown in Fig.~\ref{fig:EFig_vH}a are the lower and upper boundaries of the continuum from pairs of non-interacting triplets (equivalent to the two-magnon continuum). There,  the sharp peaks for the $\Delta S_2$ excitation appear above the upper  boundary of the continuum throughout the probed region of reciprocal space. The small positive value therefore suggests a weak \emph{repulsive} interaction between the triplets formed after a quadrupolar $\Delta S_2$ excitation~\cite{white1993prb}, and the sharpness of the peaks suggests that a bound state of the triplet pairs forms.

We here provide a semi-quantitative analysis to show that at sufficiently high energies, a pair of triplet excitations form a bound state above the two-triplet continuum in presence of weak repulsion between them. The $\Delta S_1$ dispersion can be expanded around $q_{\parallel}\simeq\pm 0.25$, 
which are the likely momentum states contributing to the formation of the highest energy triplet pair state at $q_{\parallel}\simeq0$. From Fig.~\ref{fig:EFig_vH}d, we obtain 
\begin{equation}
E(q_{\parallel}\simeq\pm 0.25+\delta q_{\parallel})\simeq \omega_{S_1}(q_{\parallel}=\pm 0.25)+\frac{\hslash^2 (\delta q_{\parallel})^2}{2m^{*}}.
\end{equation}
A Schr\"{o}dinger equation can thus be set up for these two quasiparticles resulting from independent $\Delta S_1$ excitations:
\begin{equation}
\Big[-\frac{\hslash^2}{2m^{*}}\frac{\partial^2}{\partial r_1^2}-\frac{\hslash^2}{2m^{*}}\frac{\partial^2}{\partial r_2^2}+V(r_1,r_2)\Big]\Psi(r_1,r_2)=(E-2\omega_{S_1}(q_{\parallel}\simeq0.25))\Psi(r_1,r_2).
\end{equation}
Since we are interested in the triplet pair state at $q_{\parallel}\simeq0$, we can neglect the center of mass motion ($\delta q^{\rm cm}_{\parallel}=\delta q_{\parallel}=0$ and $E=\tilde{E}+\frac{\hslash^2 (\delta q_{\parallel})^2}{2m_{cm}^{*}}=\tilde{E}$) and consider just 
the dynamics in the relative coordinate $r=r_1-r_2$:
\begin{equation}
\Big[-\frac{\hslash^2}{2m^{*}}\frac{\partial^2}{\partial r^2}+V(r)\Big]\Psi(r)=(\tilde{E}-2\omega_{S_1}(q_{\parallel}\simeq0.25))\Psi(r).
\end{equation}
Assuming a repulsive $\delta$-like potential among the $\Delta S_{tot}=2$ excitations~\cite{white1993prb}, $V(r)=\lambda\delta(r)$, with $\lambda>0$ and noting that the effective mass to be negative $m^{*}=-|m^{*}|$ we get
\begin{equation}
\Big[-\frac{\hslash^2}{2|m^{*}|}\frac{\partial^2}{\partial r^2}-\lambda\delta(r)\Big]\Psi(r)=E^{\prime}\Psi(r),
\end{equation}
where we have multiplied the equation by $-1$ and then redefined $E^{\prime}=-\tilde{E}+2\omega_{S_1}(q_{\parallel}\simeq0.25)$. Elementary quantum mechanics then predicts the existence of a bound state with energy  
\begin{equation}\label{Eq:E}
E^{\prime}_{B}=-\frac{|m^{*}|\lambda^2}{2\hslash^2}.
\end{equation}

Eq.~(\ref{Eq:E}) implies the existence of a bound state above the continuum, since for $E^{\prime}<0$ we have  $\tilde{E}>2\omega_{S_1}(q_{\parallel}\simeq0.25)$, and $\tilde{E}=2\omega_{S_1}(q_{\parallel}\simeq0.25)+|E^{\prime}_{B}|$.
From this expression, we can \emph{a posteriori} estimate the strength of the quasiparticle interaction $\lambda$ (which has units of energy times length, so we divide by the lattice unit $c$) assuming $|E^{\prime}_{B}|\simeq 7$~meV:
\begin{equation}
\lambda/c=\sqrt{4\frac{\hslash^2}{2|m^{*}|}|E^{\prime}_{B}|} \simeq 39~\text{meV} \simeq 1.6J.
\end{equation}
This is a rather weak interaction if we compare it with the $\Delta S_1$ bandwidth 
$W\simeq 2.3J$ and therefore validates our previously estimated energy scales of the triplet pair density of states and continuum boundaries.

To further confirm this estimate, we simulated a Bose Hubbard model on a 1D lattice of a finite length with open boundary conditions with an on-site attractive interaction $U_b=-|U_b|$ with $N_b=2$ bosons 
\begin{equation}
H=-t_b \sum_{i} (b^{\dag}_i b^{\phantom\dagger}_{i+1}+{\rm h.c.}) + \frac{U_{b}}{2}\sum_i n^b_i (n^b_i-1).
\end{equation}
We have assumed $t_b=0.575J$ such that the bandwidth is $W_b=4t_b=2.3J$. We find that on a long chain of $L=64$ sites, the system shows a bound state with energy $E_{B}= E(U_b=-1.675J)-E(U_b=0)\simeq-0.28J$ $\sim-6.7$ meV, consistent with our estimation above using an attractive $\delta$-like potential between the quasiparticles.

\section*{Supplementary Note 7: Computational details to reproduce the DMRG results}

Here we provide instructions on how to reproduce the DMRG results used in the main text. The results reported in this work were obtained with DMRG++ versions 6.01 and PsimagLite versions 3.01. The DMRG++ computer program~\cite{alvarez0209} can be obtained with:
\begin{verbatim}
git clone https://github.com/g1257/dmrgpp.git
git clone https://github.com/g1257/PsimagLite.git
\end{verbatim}
The main dependencies of the code are BOOST and HDF5 libraries.
To compile the program:
\begin{verbatim}
cd PsimagLite/lib; perl configure.pl; make
cd ../../dmrgpp/src; perl configure.pl; make
\end{verbatim}

The DMRG++ documentation  can  be  found  at  \verb! https://g1257.github.io/dmrgPlusPlus/manual.html! or  can  be  obtained  by doing 
\verb!cd dmrgpp/doc; make manual.pdf!. In the description of the DMRG++ inputs below,
we follow very closely the description in the supplemental material of Ref.~\cite{scheie2021witnessing}, where similar calculations were performed.
%We have also provided inputs and scripts to reproduce our figures at \url{Insert Url.}
Other inputs and scripts can be made available upon request. \\

\noindent\textbf{a) Obtaining zero-temperature spectra}

The $T=0$ results can be reproduced as follows. We first run 
\verb!./dmrg -f inputGS.ain -p 12! to obtain the ground state wave-function and ground state energy with 12 digit precision using the \verb!-p 12! option. The \verb!inputGS.ain! has the form
\begin{verbatim}
##Ainur1.0
TotalNumberOfSites=100;
NumberOfTerms=2;

### 1/2(S^+S^- + S^-S^+) 
gt0:DegreesOfFreedom=1;
gt0:GeometryKind="chain";
gt0:GeometryOptions="ConstantValues";
gt0:dir0:Connectors=[1.0];

### S^zS^z part
gt1:DegreesOfFreedom=1;
gt1:GeometryKind="chain";
gt1:GeometryOptions="ConstantValues";
gt1:dir0:Connectors=[1.0];

Model="Heisenberg";
HeisenbergTwiceS=2;
SolverOptions="twositedmrg";
InfiniteLoopKeptStates=100;
FiniteLoops=[[49, 1000, 0],
[-98, 1000, 0],
[98, 1000, 0],
[-98, 1000, 0],
[98, 1000, 0]];

# Keep a maximum of 1000 states, but allow SVD truncation with 
# tolerance 1e-10 and minimum states equal to 100
TruncationTolerance="1e-10,100";
# Symmetry sector for ground state S^z_tot=0
TargetSzPlusConst=100    
\end{verbatim}

Here we showed the input for S=1 (note \verb!HeisenbergTwiceS=2!). The parameter TargetSzPlusConst should be equal $Sz+L$, 
where $Sz$ is the targeted $Sz$ sector and $L$ is the system size. 
The next step is to calculate dynamics for the $S^0(q_{\parallel},\omega)$, $S^1(q_{\parallel},\omega)$ and $S^2(q_{\parallel},\omega)$ spectral functions using the saved ground state as an input. For simplicity, we discuss the $S^1(q_{\parallel},\omega)$ first. It is convenient to do the dynamics run in a subdirectory \verb!S1!, so \verb!cp inputGS.ain S1/inputS1.ado! and add/modify the following lines in \verb!inputS1.ado!
\begin{verbatim}
SolverOptions="twositedmrg,restart,minimizeDisk,CorrectionVectorTargeting";
# The finite loops now start from the final loop of the gs calculation
FiniteLoops=[[-98, 2000, 2],
[98, 2000, 2],
[-98, 2000, 2],
[98, 2000, 2]];
TruncationTolerance="1e-7,100";

# RestartFilename is the name of the GS .hd5 file (extension is not needed) 
RestartFilename="../inputGS";

# The weight of the g.s. in the density matrix
GsWeight=0.1;
# Legacy, set to 0
CorrectionA=0;
# Fermion spectra has sign changes in denominator. 
# For boson operators (as in here) set it to 0
DynamicDmrgType=0;
# The site(s) where to apply the operator below. Here it is the center site.
TSPSites=[49];
# The delay in loop units before applying the operator. Set to 1 
TSPLoops=[1];
# If more than one operator is to be applied, how they should be combined.
# Irrelevant if only one operator is applied, as is the case here.
TSPProductOrSum="sum";
# How the operator to be applied will be specified
string TSPOp0:TSPOperator=expression;
# The operator expression
string TSPOp0:OperatorExpression="splus";
# How is the freq. given in the denominator (Matsubara is the other option)
CorrectionVectorFreqType="Real";
# This is a dollarized input, so the 
# omega will change from input to input.
CorrectionVectorOmega=$omega;
# The broadening for the spectrum in omega + i*eta
CorrectionVectorEta=0.25;
# The algorithm
CorrectionVectorAlgorithm="Krylov";
#The labels below are ONLY read by manyOmegas.pl script
# How many inputs files to create
#OmegaTotal=600
# Which one is the first omega value
#OmegaBegin=-2.0
# Which is the "step" in omega
#OmegaStep=0.025
# Because the script will also be creating the batches, 
# indicate what to measure in the batches
#Observable=sminus    
\end{verbatim}

Then  all  individual  inputs  (one  per $\omega$ in  the  correction  vector  approach)  can  be  generated  and  submitted  using  the \verb!manyOmegas.pl! script which can be found in the \verb!dmrgpp/src/script! folder:
\begin{verbatim}
perl manyOmegas.pl inputS1.ado BatchTemplate.pbs <test/submit>.
\end{verbatim}
It is recommended to run with \verb!test! first to verify correctness, before running with \verb!submit!.  Depending on the machine and scheduler, the \verb!BatchTemplate! can be e.g. a PBS or SLURM script. 
The key is that it contains a line \verb!./dmrg -f $$input "<X0|$$obs|P2>" -p 12! which allows \verb!manyOmegas.pl! to fill in the appropriate input for each generated job batch. 
After all outputs have been generated,
\begin{verbatim}
perl procOmegas.pl -f inputS1.ado -p
perl pgfplot.pl
\end{verbatim}
can be used to process and plot the results (these scripts are also given in 
\verb!dmrgpp/src/script! folder).

For the calculation of the quadrupolar spectral function $S^2(q_{\parallel},\omega)$ 
we would have to substitute in the file above the lines
\begin{verbatim}
OperatorExpression="splus*splus";
#Observable=sminus*sminus    
\end{verbatim}
These lines will allow to apply the operator $(S^{+}_c)^2$ at the center of the chain and
to measure its hermitian conjugate $(S^{-}_j)^2$ on all others sites when we run with \verb!./dmrg -f $$input "<X0|$$obs|P2>"!.

For the calculations of the $S^0(q_{\parallel},\omega)$ we would need to perform four separate calculations:  
\begin{align}
    S^0_{\rm part-1}(q_{\parallel},\omega)&=\frac{1}{L}\sum^{L}_{j=1}e^{iq_{\parallel}(j-c)}\Big\{-\frac{1}{\pi}{\rm Im}\Big[\langle\psi|S^{z}_{j}S^{z}_{j+1}\frac{1}{\omega-H+i\eta}S^{z}_{c}S^{z}_{c+1}|\psi\rangle\Big]\Big\}\\
    S^0_{\rm part-2}(q_{\parallel},\omega)&=\frac{1}{L}\sum^{L}_{j=1}e^{iq_{\parallel}(j-c)}\Big\{-\frac{1}{\pi}{\rm Im}\Big[\langle\psi|(S^{+}_{j}S^{-}_{j+1}+S^{-}_{j}S^{+}_{j+1})\frac{1}{\omega-H+i\eta}S^{z}_{c}S^{z}_{c+1}|\psi\rangle\Big]\Big\}\\    
    S^0_{\rm part-3}(q_{\parallel},\omega)&=\frac{1}{L}\sum^{L}_{j=1}e^{iq_{\parallel}(j-c)}\Big\{-\frac{1}{2\pi}{\rm Im}\Big[\langle\psi|(S^{+}_{j}S^{-}_{j+1}+S^{-}_{j}S^{+}_{j+1})\frac{1}{\omega-H+i\eta}S^{+}_{c}S^{-}_{c+1}|\psi\rangle\Big]\Big\}\\
    S^0_{\rm part-4}(q_{\parallel},\omega)&=\frac{1}{L}\sum^{L}_{j=1}e^{iq_{\parallel}(j-c)}\Big\{-\frac{1}{2\pi}{\rm Im}\Big[\langle\psi|(S^{+}_{j}S^{-}_{j+1}+S^{-}_{j}S^{+}_{j+1})\frac{1}{\omega-H+i\eta}S^{-}_{c}S^{+}_{c+1}|\psi\rangle\Big]\Big\}    
\end{align}
We should then combine the results as
$S^0(q_{\parallel},\omega)=S^0_{\rm part-1}(q_{\parallel},\omega)+S^0_{\rm part-2}(q_{\parallel},\omega)+\frac{1}{2}[S^0_{\rm part-3}(q_{\parallel},\omega)+S^0_{\rm part-4}(q_{\parallel},\omega)]$. For simplicity, we here only provide instructions for the calculation of the spectral function $S^0_{\rm part-1}(q_{\parallel},\omega)$. This needs the following modifications in the \verb!inputS0part1.ado!
\begin{verbatim}
# minimizeDisk option should be removed
SolverOptions="twositedmrg,restart,CorrectionVectorTargeting";
# The finite loops need to have save option 3 at the last loop.
# This is to prepare observation of two-point correlations
FiniteLoops=[[-98, 2000, 2],
[98, 2000, 2],
[-98, 2000, 2],
[98, 2000, 3]];
string TruncationTolerance="1e-7,100";

# The sites where to apply the operator S^{z}_{c}S^{z}_{c+1}. 
# Here center and center+1 sites
TSPSites=[49,50];
# The delay in loop units before applying the operator. Set to 0
TSPLoops=[0,0];
TSPProductOrSum="product";
# How the operator to be applied will be specified
string TSPOp0:TSPOperator=expression;
string TSPOp0:OperatorExpression="sz";
string TSPOp1:TSPOperator=expression;
string TSPOp1:OperatorExpression="sz";
\end{verbatim}
In this case, the \verb!BatchTemplate! needs to contain the lines running \verb!dmrg! and the \verb!observe! executables 
\begin{verbatim}
    ./dmrg -f $$input  -p 12
    ./observe -f $$input -p 12 "<gs|sz;sz|P2>"
\end{verbatim} 
Straightforward modifications of the script \verb!manyOmegas.pl! need to performed 
to gather the resulting real space correlations and their Fourier Transform. 
Input files for the calculation of $S^0_{\rm part-2}(q_{\parallel},\omega)$, $S^0_{\rm part-3}(q_{\parallel},\omega)$ 
and $S^0_{\rm part-4}(q_{\parallel},\omega)$ and associated scripts for processing 
the numerical results can be provided upon request.\\

\noindent\textbf{b) Obtaining finite-temperature spectra}

As noted in the Methods section in the main text, finite temperature $T>0$ 
calculations proceed in three steps (see also Refs.~\cite{Feiguin2005,Tiegel2013}):  We first prepare the system in a $T \rightarrow \infty$ (inverse temperature $\beta=1/T= 0$) state using a fictitious ``entangler'' Hamiltonian $H_E$ acting in an enlarged Hilbert space. In the purification approach, the geometry of the system can be described as a spin ladder, with physical sites on one leg (with even sites 0,2,4,...) and ancilla sites on the other (with odd sites 1,3,5,...). 

The entangler Hamiltonian is chosen such that its ground state $|\psi_\infty \rangle$ corresponds to the $T\rightarrow \infty$ state  of the physical system when the ancilla degrees of freedom are traced out, $\rho={\rm Tr}\left[|\psi_\infty\rangle\langle\psi_\infty|\right]_A$. We use a conventional entangler Hamiltonian $H_E$ conserving only the total magnetization of the enlarged physical+ancilla system.  

We find the ground state of $H_E$ by running \verb!./dmrg -f Entangler.ain -p 12!, where \verb!Entangler.ain! for L=32 sites (32 physical and 32 ancillas) is given by

\begin{verbatim}
##Ainur1.0
TotalNumberOfSites=64;
NumberOfTerms=2;
gt0:DegreesOfFreedom=1;
gt0:GeometryKind="ladder";
gt0:LadderLeg=2;
gt0:GeometryOptions="ConstantValues";
gt0:dir0:Connectors=[0.0];
gt0:dir1:Connectors=[-1.0];

gt1:DegreesOfFreedom=1;
gt1:GeometryKind="chain";
gt1:GeometryOptions="ConstantValues";
gt1:dir0:Connectors=[0];

Model="Heisenberg";
integer HeisenbergTwiceS=2;
SolverOptions="twositedmrg,MatrixVectorOnTheFly";
InfiniteLoopKeptStates=100;
FiniteLoops=[[ 31, 1000, 0],
[-62, 1000, 0],
[ 62, 1000, 0],
[-62, 1000, 0]];
# Keep a maximum of 1000 states, but allow 
# density matrix truncation with tolerance and minimum states as below
TruncationTolerance="1e-8,100";
TargetSzPlusConst=64
\end{verbatim}

Second, the system is cooled through evolving in imaginary time with the physical Hamiltonian $H$ acting only on  physical sites $e^{-\beta(H\otimes I)/2}$, where $I$ is the identity operator in the ancilla space.  To compute the thermal state of the system $|\psi(\beta)\rangle={e^{-\beta(H\otimes I)/2}}|\psi_\infty\rangle$, we used the Krylov algorithm for time evolution with an imaginary time step $\Delta\beta=0.025/J$. The time evolution is done with an evolution operator $exp[-\beta^{\prime} H /2]$ where we define $\beta^{\prime}=\beta/2$ in units of $1/J$. 

Technically, we compute the imaginary time evolution with \verb!./dmrg -f EvolutionT.ain -p 12!, where \verb!EvolutionT.ain! reads (we report only the main differences with respect to the previous input)
\begin{verbatim}
gt0:GeometryKind="ladder";
gt0:LadderLeg=2;
gt0:GeometryOptions="none";
gt0:dir0:Connectors=[1.0, 0.0, 1.0, 0.0, 1.0, 0.0, 1.0, 0.0, ...];
gt0:dir1:Connectors=[0.0, 0.0, 0.0, 0.0, ...];

gt1:GeometryKind="ladder";
gt1:LadderLeg=2;
gt1:GeometryOptions="none";
gt1:dir0:Connectors=[1.0, 0.0, 1.0, 0.0, 1.0, 0.0, 1.0, 0.0, ...];
gt1:dir1:Connectors=[0.0, 0.0, 0.0, 0.0, ...];

string PrintHamiltonianAverage="s==c";
string RecoverySave="@M=100,@keep,1==1";
SolverOptions="twositedmrg,restart,TargetingAncilla";

# Notice the save option 3 at the last loop corresponding to the 
# physical temperature of interest (in this specific case 250K)
# This allows to measure static correlation functions
FiniteLoops=[[ 62, 2000, 2],[-62, 2000, 2],[ 62, 2000, 2],[-62, 2000, 2],
[ 62, 2000, 2],[-62, 2000, 2],[ 62, 2000, 2],[-62, 2000, 2],
[ 62, 2000, 2],[-62, 2000, 2],[ 62, 2000, 2],[-62, 2000, 2],
[ 62, 2000, 2],[-62, 2000, 2],[ 62, 2000, 2],[-62, 2000, 2],
[ 62, 2000, 2],[-62, 2000, 2],[ 62, 2000, 2],[-62, 2000, 2],
[ 62, 2000, 2],[-62, 2000, 2],[ 62, 2000, 2],[-62, 2000, 3]];
RestartFilename="Entangler";
TSPTau=0.025;
TSPTimeSteps=5;
TSPAdvanceEach=62;
TSPAlgorithm="Krylov";
TSPSites=[1];
TSPLoops=[0];
TSPProductOrSum="sum";
GsWeight=0.1;
string TSPOp0:TSPOperator=expression;
string TSPOp0:OperatorExpression="identity";

\end{verbatim}
Above, the \verb!gt1:dir0! connectors give the interactions couplings along the leg
direction of the two-leg ladder geometry, so the alternating pattern \verb![1.0, 0.0, 1.0, 0.0,...]! just means that we evolve in imaginary time the physical sites rather than the ancillas. The imaginary time $\beta^{\prime}$ can be obtained with\\ 
\verb!h5dump -d /Def/FinalPsi/TimeSerializer/Time <hd5>!, 
where \verb!<hd5>! should be replaced with the name of the \verb!hd5! file of interest.
The targeted $\beta^{\prime}$ value is given by $\beta^{\prime}(J,T)=1/(2JT)$, where J and T are given in Kelvin. Finally, the arguments to \verb!RecoverySave! mean that we keep a maximum of 100 \verb!hd5! outputs, and output one in every loop (when the condition \verb!1==1! holds). 
We also observe dipolar as well as quadrupolar correlation functions at finite temperature
to extract their characteristic correlation lengths. This is done by doing 
\begin{verbatim}
    ./observe -f EvolutionT.ain -p 12 "<time|sz;sz|time>"
    ./observe -f EvolutionT.ain -p 12 "<time|splus*splus;sminus*sminus|time>"
\end{verbatim}

Finally, the dynamical spin structure factor is calculated using the operator ${\cal L}=H\otimes I+I\otimes(-H)$ with the real space expression in Eq.~\ref{eq:Liouville}.
For the DMRG++ inputs, the dynamics calculation proceeds similarly to the $T=0$ case, but with number of sites and precision as in the preceding $T>0$ step. 
We do, however, need to additionally add/modify the following lines
\begin{verbatim}
gt0:dir0:Connectors=[1.0, -1.0, 1.0, -1.0, 1.0, -1.0, 1.0, -1.0, ...];
gt0:dir1:Connectors=[0.0, 0.0, 0.0, 0.0, ...];
gt1:dir0:Connectors=[1.0, -1.0, 1.0, -1.0, 1.0, -1.0, 1.0, -1.0, ...];
gt1:dir1:Connectors=[0.0, 0.0, 0.0, 0.0, ...];
RestartFilename="Recovery23EvolutionT.hd5";
SolverOptions="CorrectionVectorTargeting,restart,twositedmrg,minimizeDisk,fixLegacyBugs";
integer RestartSourceTvForPsi=0;
vector RestartMappingTvs=[-1, -1, -1, -1];
integer RestartMapStages=0;
# This is the structure of FiniteLoops for dynamics
FiniteLoops=[[-98, 2000, 2],[98, 2000, 2],[-98, 2000, 2],[98, 2000, 2]];
string TruncationTolerance="1e-7,100";
\end{verbatim}
Above, the restart filename should be chosen to match the \verb!hd5! file of interest. All rung couplings are zero. As before, we have abbreviated the arrays of coupling constants.

\bibliography{references2}